%% file: main.tex
\documentclass[11pt]{article}
\usepackage[margin=1in]{geometry}

\usepackage[
backend=biber,
style=ieee,
citestyle=numeric-comp,
maxbibnames=8,
minbibnames=4,
sorting=none,
url=false,
doi=true,
isbn=false,
uniquename=false,
alldates=year,
clearlang=true,]{biblatex}
\DeclareFieldFormat[article,unpublished,misc,preprint]{title}{``#1''}

\let\oldcite\cite
\renewcommand{\cite}[1]{{\mbox{\oldcite{#1}}}}

\AtEveryBibitem{\clearlist{publisher}}
\AtEveryBibitem{\clearfield{note}}
\AtEveryBibitem{\clearlist{language}}
\AtEveryBibitem{\clearname{editor}}

\usepackage{booktabs}
\usepackage{siunitx}
\usepackage{fancyhdr}
\usepackage{hyperref}
\usepackage[acronym]{glossaries}
\usepackage[font=small,labelfont=it]{caption}
\usepackage{graphicx}
\usepackage{subcaption}
\usepackage[dvipsnames, usenames]{xcolor}
\usepackage{soul}
\usepackage{float}
\usepackage{tabularx,array,amssymb,arydshln}
\usepackage{xcolor}
\usepackage{threeparttable}
\usepackage{booktabs}
\usepackage{tabularx}
\usepackage{array}
\usepackage[table]{xcolor} % <-- IMPORTANT for \rowcolor and \cellcolor
\usepackage{titlesec}
\usepackage{xfrac}
\usepackage{dblfloatfix}
\usepackage{siunitx}
\usepackage{arydshln}
\usepackage{subcaption} % in your preamble
\usepackage{booktabs}

\usepackage{makecell}
\usepackage{multirow}
\renewcommand\arraystretch{1.4} % Adjust row height
\usepackage{ragged2e}
\usepackage{placeins}
\usepackage{cancel}

\usepackage{cprotect}
\usepackage{amsmath,amsfonts,amssymb, bm, bbm}

\DeclareMathOperator{\He}{He}

\renewcommand{\vec}[1]{\mathbf{\bm{#1}}}

\newcommand{\appropto}{\mathrel{\vcenter{
  \offinterlineskip\halign{\hfil$##$\cr
    \propto\cr\noalign{\kern2pt}\sim\cr\noalign{\kern-2pt}}}}}

\usepackage{mathtools}

\usepackage{lipsum}
\input{glossary}

\begin{document}

% \title{maya
\title{Online gas classification at the edge}
\author{
\textbf{Ton Juny Pina}\textsuperscript{1,2,*},\quad
\textbf{Hugh Greatorex}\textsuperscript{1,2,3},\quad
\textbf{Elisabetta Chicca}\textsuperscript{1,2}
}
\maketitle

\begingroup
\renewcommand{\thefootnote}{}
\footnotetext{\raggedright
\textsuperscript{1} Bio-Inspired Circuits and Systems (BICS) Lab, Zernike Institute for Advanced Materials, University of Groningen, Netherlands.\\
\textsuperscript{2} Groningen Cognitive Systems and Materials Center (CogniGron), University of Groningen, Netherlands.\\
\textsuperscript{3} Institute of Neuroinformatics, University of Zurich and ETH Zurich.\\
\textsuperscript{*} Corresponding author: t.juny.pina@rug.nl}
\endgroup
\maketitle

\thispagestyle{plain}
\pagestyle{plain}

\begin{abstract}

Field applications in chemical sensing, from portable safety monitors to mobile robots, require models that are lightweight, fast to respond, and can be recalibrated from limited labelled data.
Current gas classification pipelines assume controlled sampling cycles and offline feature extraction, and rely on computationally expensive models.
With a few thousand neurons and sub-milliwatt power consumption, insects reliably classify odours in open environments.
Inspired by the insect olfactory system, we present a neural network architecture that classifies raw sensor array readings using a sparse, high-dimensional binary representation. 
We evaluate it on random sequences of single gases and binary mixtures, introducing new samples dynamically and classifying gas identity from the transient sensor response. 
In this setting, the architecture out-performs a linear \gls{svm} on gas mixtures and matches an optimised non-linear \gls{svm}, while maintaining online class-level frequency estimates that distinguish familiar from novel stimuli.
It relies exclusively on binary weights and Hebbian-based local learning, remaining computationally lightweight and well suited to ultra-low-power neuromorphic processors.
We demonstrated the network on-chip with a hazardous gas detection experiment running on the mixed-signal neuromorphic chip TEXEL. 
The on-chip classifier yielded an accuracy of 74.6\% while running at 482 $\unit{\nano\joule}$ per inference and 27 \unit{\micro\watt} of static-power.

\vspace{1em}

\end{abstract}

\section{Introduction}
\label{sec:introduction}

% [Applications of online odour classification and its requirements]
Gas sensing in real-world environments presents a challenging classification problem.
Turbulent plumes exhibit complex spatial and temporal structure, so traces of the same mixture can differ substantially with changes in airflow and local concentration gradients~\cite{connor_quantification_2018, kadakia_odour_2022}. 
The sensors themselves drift in sensitivity and selectivity: the same sensor can produce different responses to an identical mixture over its operating life~\cite{vergara_chemical_2012, dennler_drift_2022}, so deployed devices require periodic recalibration.
Established classification models are poorly matched to these conditions~\cite{zhang_new_2022, peng_gas_2018, wei_development_2019, wang_exploiting_2023, du_vision_2024, kwon_efficient_2021, kang_high_2022, xiong_gas_2021, vanarse_application_2022, huo_bio-inspired_2023, xue_novel_2024}: they assume controlled sampling cycles, employ offline feature extraction, can issue predictions only after complete cycles and rely on computationally expensive operations.
To operate autonomously and for extended durations -- in environmental monitoring, industrial safety and mobile robotics~\cite{sekhar_chemical_2010, kim_development_2018, lu_electronic_2022, jain_review_2025} -- devices must instead be compact, energy-efficient and computationally lightweight, and able to recalibrate on-device from few examples.
% This is currently solved using electrochemical sensors with very selective responses, and detection is based on threshold comparison. Electrochemical sensors are very selective but have drawbacks: short sensor lifetime, big form factor.
% The incorporation of new sensor technologies and form factors, such as \gls{mems}-based sensors, requires 

% [Bio-inspired processing -> The expand-and-sparsify architecture]
Animals have evolved to operate efficiently under highly restrictive constraints on power-consumption, computational speed and available training examples. 
% With only a few thousands of neurons, the insect olfactory system allows to navigate turbulent environments, find food sources, locate their nest and communicate with peers at sub-mW consumption.
With only a few thousands of neurons and a sub-mW consumption~\cite{neville_novel_2018, panda_direct_2025}, the insect olfactory system outperforms electronic-nose systems in selectivity, sensitivity, efficiency and latency, providing a great example for efficient olfactory processing.
This success arises from highly specialized receptors and efficient processing units.
On the animal's head, olfactory receptor neurons innervate hair-like structures called sensilla, interacting with odorants in the air.
The olfactory receptor neurons transmit the sensory information into the \gls{al}, the first processing centre in the olfactory pathway.  
Dense receptor signals from the \gls{al} are expanded into a sparse, \gls{hd} odour representation in the \gls{mb}~\cite{galizia_olfactory_2014}. 
This \textit{expand-and-sparsify} network induces a locality-preserving \gls{hd} embedding suited to associative learning, allowing the animal rapid and robust recognition~\cite{huerta_learning_2004, babadi_sparseness_2014, webb_beyond_2024}.
The computational principles of this transformation are well-studied and have been formalized as \gls{lsh}~\cite{dasgupta_neural_2017,indyk_approximate_1998, wang_hashing_2014}
% , and applied to similarity search applications~\cite{dasgupta_neural_2017, dasgupta_neural_2022}
.

% [Neuromorphic systems -> e&s architecture is hardware friendly]
Neuromorphic hardware has made significant progress in capturing the efficiency of biological systems through distributed in-memory computing, asynchronous event-based communication and local learning rules~\cite{mead_neuromorphic_1990, indiveri_vlsi_2006}. 
These advantages typically come at the cost of low-resolution weights, noisy neurons, and limited learning methods available.
With sparse binary weights, sparse \gls{hd} binary representations and associative learning, the expand-and-sparsify network supports efficient odour classification on \gls{snn} processors with on-chip learning~\cite{moradi_scalable_2018, greatorex_neuromorphic_2025, orchard_efficient_2021}. 
% Efficient on-chip computation of the sparse \gls{hd} projection has been demonstrated in the RISC-V platform PULPv3 ($2 \unit{\milli\watt}$)~\cite{montagna_pulp-hd_2018}, and the read-out has been demonstrated on the mixed-signal neuromorphic chip TEXEL ($27 \unit{\micro\watt}$ static power)~\cite{greatorex_neuromorphic_2025}.

% [What did we do]
Taking inspiration from the olfactory processing in the insect brain, we deploy the expand-and-sparsify network in a dynamic and online gas classification framework.
Despite relying exclusively on binary weights and local Hebbian-based learning, the architecture outperforms linear classification on gas mixtures and yields a feature-space linearization comparable to non-linear kernel methods, classifying gas identity from the transient sensor response rather than a settled steady state.
The network is highly data-efficient, reaching near-peak accuracy from only a few tens of labelled exposures, and we further show that the same architecture maintains online class-frequency estimates, distinguishing familiar from novel stimuli.
Finally, we demonstrate the network on-chip, running a hazardous-gas detection experiment on the mixed-signal neuromorphic processor TEXEL at 482 $\unit{\nano\joule}$ per inference and roughly 27 $\unit{\micro\watt}$ of static power, establishing the architecture's suitability for ultra-low-power edge deployment.

\section{Results}

\subsection{A fly-inspired system for online gas classification}
\label{sec:system_intro}

Field gas sensing presents a challenging classification problem. Turbulent transport produces intermittent odour plumes, so a single source yields variable traces over time. 
Stimuli are commonly composed of many volatiles that cannot be solved by a single sensor, and many sensor technologies are inherently non-selective.
A deployed system must therefore infer the gas identity from a global, dynamic response of the sensor array rather than from a static fingerprint.
The sensors also drift in sensitivity and selectivity over their operating life, requiring systems to periodically recalibrate~\cite{vergara_chemical_2012, dennler_drift_2022}.

\begin{figure}[h]
    \centering
    \includegraphics[width=\linewidth]{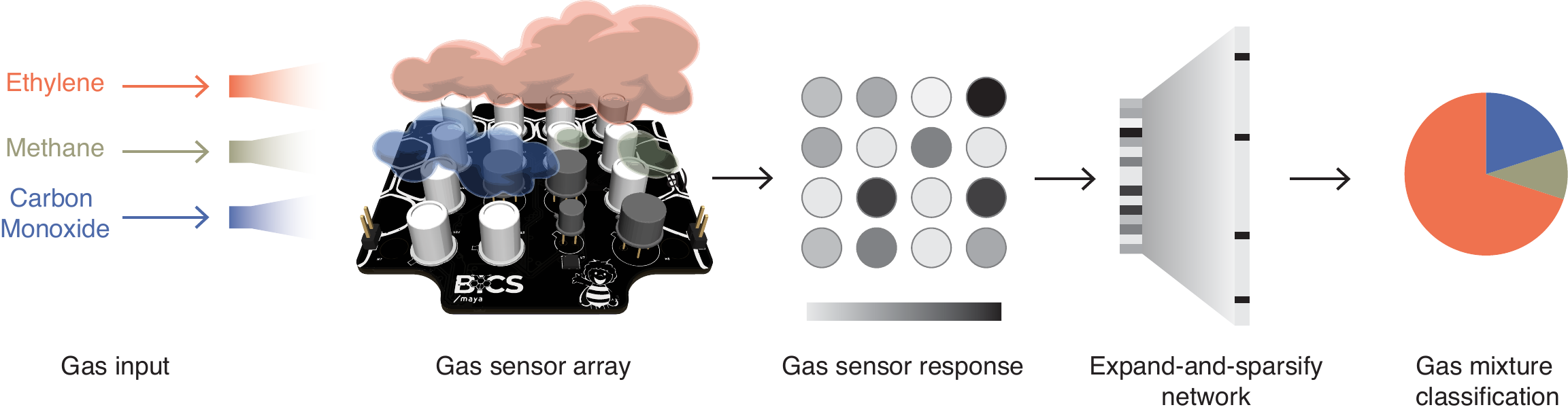}
    \caption{\textbf{System overview.
    }
    The sensor array is exposed to a randomized sequence of gas mixtures. Its response is then processed by the expand-and-sparsify network producing online category estimates.
    }
    \label{fig:main}
\end{figure}

To study classification under these conditions we built a sensing platform around an array of commercial \gls{mox} sensors on a custom board (Fig.~\ref{fig:main}; Methods).
We chose \gls{mox} because they are inexpensive, long-lived and compatible with \gls{mems} fabrication, at the cost of poor per-gas selectivity — a trade-off that deliberately shifts the discrimination task from the sensor chemistry onto the downstream processing.
Gases are delivered by three independent \glspl{mfc}, allowing us to arbitrarily sample single analytes (ethylene, methane and CO) and their mixtures. 
Instead of the conventional baseline–exposure–recovery cycle~\cite{zhang_new_2022, peng_gas_2018, wei_development_2019, wang_exploiting_2023, du_vision_2024, kwon_efficient_2021, kang_high_2022, xiong_gas_2021, vanarse_application_2022, huo_bio-inspired_2023, xue_novel_2024}, we present mixtures sequentially in windows shorter than the characteristic response times of the sensors, so the array does not settle to steady state. 
The initial condition of every exposure is therefore set by the preceding stimulus, producing a broad distribution of transient trajectories per class, reproducing the non-stationarity of open-air sensing.

\begin{figure}[h]
    \centering
    \includegraphics[width=\linewidth]{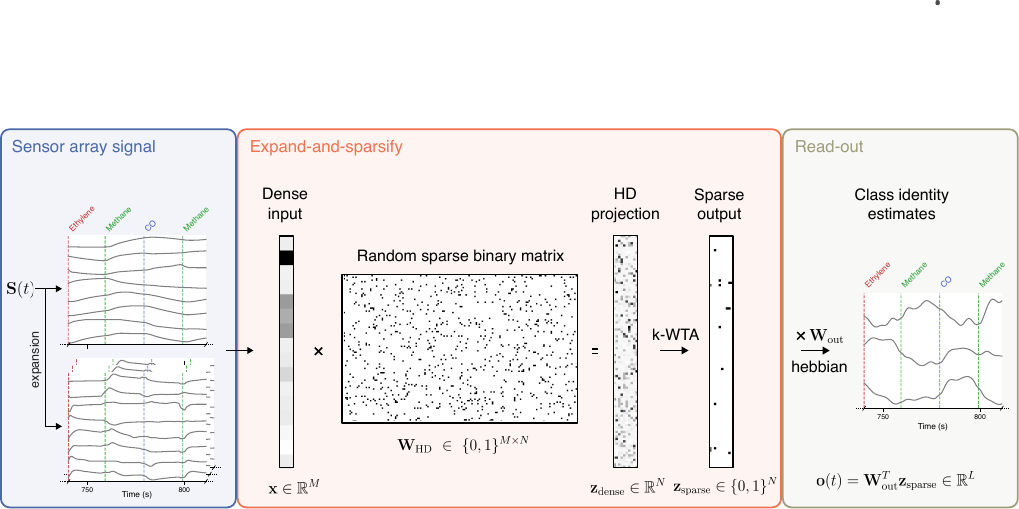}
    \vspace{5pt}
    \caption{\textbf{The expand-and-sparsify architecture.
    }
    As the sensor array responds to a randomized sequence of gas mixtures, the first and second order derivatives of each sensor signal are approximated using backward differences, and the 56 pairwise channel ratios and differences are appended to form a 136-dimensional feature vector.
    This signal is then projected onto a higher-dimensional space via multiplication with a random sparse binary matrix, and subsequently sparsified through a $k$-WTA mechanism. 
    A linear read-out layer produces online class-identity estimates.
    }
    \label{fig:introductory}
\end{figure}

At the core of our pipeline is the expand-and-sparsify network. It works in two steps: the sensor reading is mapped onto a high-dimensional representation through a fixed random binary matrix, and a k-winner-take-all step then keeps only the most active units, producing a sparse, high-dimensional binary code (Fig.~\ref{fig:introductory}). 
A single feed-forward layer then maps this code to the class outputs.
The key property of the expand-and-sparsify network is that it induces a locality-preserving embedding: neighbouring inputs map to overlapping sparse codes while long-range relationships are attenuated, realising an explicit and computationally efficient feature map analogous to those constructed implicitly in \gls{svm} through a kernel~\cite{smola_tutorial_2004}.
In this way, non-linear separability is achieved with sparse binary projections, keeping the network computationally light-weight and compatible with the constraints of ultra-low-power neuromorphic processors~\cite{greatorex_neuromorphic_2025, davies_loihi_2018}.
The fidelity of the induced kernel depends on the input dimensionality, so the sensor readings are expanded into 136-dimensional feature vectors with the first- and second-order temporal derivatives (approximated online through backward differences), and the 56 pairwise channel differences and 56 ratios.

\subsection{Online classification of dynamic mixtures}
\label{sec:classification}

\begin{figure}
    \centering
    \includegraphics[width=\linewidth]{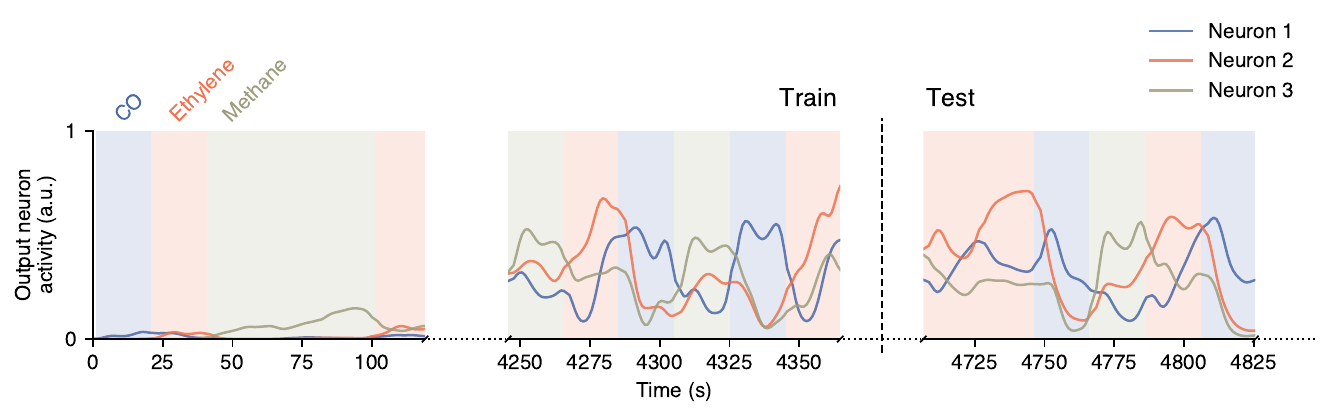}
    % \vspace{5pt}
    \caption{\textbf{Class identity estimates from the read-out neurons activity.
    }
    The activity of the read-out layer neurons as the sensor array signals are processed through the expand-and-sparsify architecture. The sensor array is exposed to a random sequence of odours (CO, ethylene, methane). During the training phase, learning signals trigger the potentiation of the output synapses, enforcing the output neurons to learn the sparse representations of its assigned category. The weights are then frozen during the testing phase.
    }
    \label{fig:readout_activity}
\end{figure}

% [Result]
We evaluated the system performance on random sequences of 600 gas presentations for training and 300 for testing.
Two datasets were considered: a simpler variant containing only single gases and a more challenging variant that also includes binary mixtures of these gases.
Gas mixtures were introduced in periods of $20\unit{\second}$, matching the characteristic sensor response times of $20 - 40\unit{\second}$ (Fig.~\ref{fig:extended_sensor_characterization}).
Performance metrics are reported in Table~\ref{tab:classification}, and Fig.~\ref{fig:readout_activity} shows the read-out neurons activity during an example training and testing sequences. 

Despite relying on binary weights trained with Hebbian learning, the
expand-and-sparsify architecture was competitive with the linear SVM, and its
advantage tracked task difficulty. On the single-gas dataset, where the classes are close to linearly separable, the 1-bit Hebbian read-out trailed the linear \gls{svm} ($0.881 \pm 0.014$ vs $0.923$), whereas on the more challenging binary-mixture dataset it outperformed it ($0.645 \pm 0.017$ vs $0.563$). This reversal is consistent with the sparse \gls{hd} embedding inducing a non-linear kernel: the benefit is largest precisely where linear separation fails, demonstrating online non-linear classification on a continuous stream of sensor data.

To assess the quality of the sparse HD representation independently of the read-out constraints, we included a 64-bit gradient-descended linear read-out as an upper-bound reference. Evaluated on the same sequential split as the baselines, it was competitive with the optimised non-linear \gls{svm} on both datasets: on the single-gas dataset it slightly exceeded the \gls{rbf} \gls{svm} ($0.952 \pm 0.003$ vs $0.940$), and on the binary-mixture dataset the two were within one accuracy point of each other ($0.791 \pm 0.008$ vs $0.800$). The kernel induced by the sparse HD embedding is therefore competitive with an optimised \gls{rbf} kernel even when class boundaries are complex, rather than being clearly surpassed by it.

\subsection{Few-shot learning}
\label{sec:few_shot_learning}

\begin{figure}
    \centering
    \includegraphics[width=0.5\linewidth]{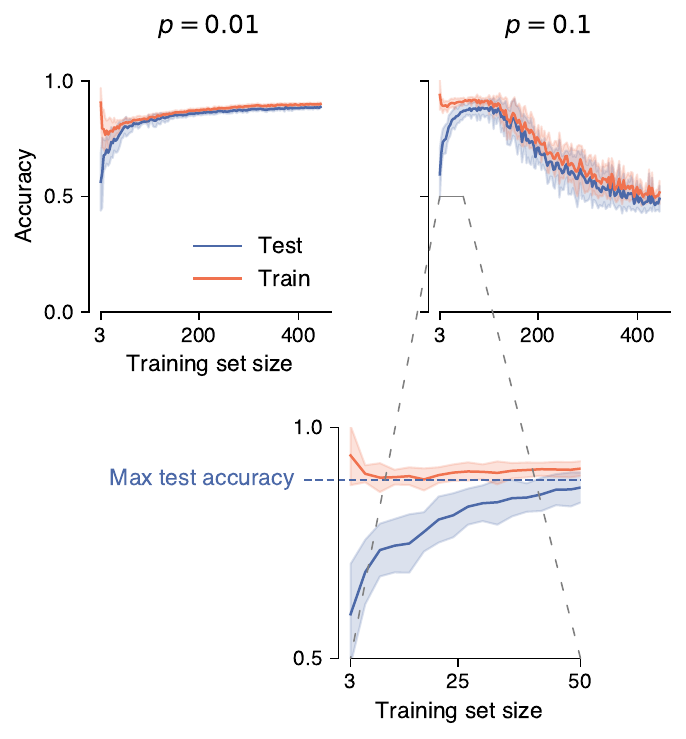}
    \caption{\textbf{Few-shot learning.
    }
    Top row: Train and test accuracy (mean $\pm$ s.d. over 10 repetitions) versus the training set size for the single-gas dataset. With p = 0.01 (left), accuracy rises gradually and peaks at 0.887 ± 0.008 with 375 exposures. A higher rate (p = 0.1, right) reaches a comparable peak (0.886 ± 0.033) with only 105 exposures, at the cost of larger variance and a decline at larger training set sizes.
    Bottom: zoomed-in view of the $p=0.1$ panel. The network reaches near-peak accuracy within the first 50 exposures. 
    }
    \label{fig:few_shot_learning}
\end{figure}

Recalibrating deployed devices is a central practical constraint in gas sensing.
Here we analysed how efficiently the network learned the category prototypes, measuring accuracy as a function of the number of training exposures (Fig.~\ref{fig:few_shot_learning}).
Learning speed was determined by the potentiation probability of the read-out weights $p$. 
With $p = 0.01$ accuracy rose gradually, peaking at $0.887 \pm 0.008$ after 375 exposures.
Raising it to $p = 0.1$ reached a comparable peak of $0.886 \pm 0.033$ after only 105.
This faster learning came at the cost of higher trial-to-trial variance, as the quality of the learnt prototypes becomes more sensitive to how representative the training subset was of the true category prototype.
At $p=0.1$, a pronounced accuracy decline appeared at large training set sizes, consistent with destructive interference between stored patterns:
as potentiation accumulated, the prototypes learnt by different read-out neurons increasingly overlapped and more samples were misclassified. The slower potentiation at $p = 0.01$ kept the weights below this regime over the range tested. 
This points to the potential benefit of a stop-learning mechanism;
alternatively, adding a weight depression term to the learning rule could mitigate the interference by reducing the overlap between stored prototypes in a continual-learning setup.

The architecture was highly data efficient, making it well suited to deployments where labelled data is scarce. With the smallest training set of 3 exposures (one per category) the network already reached a test accuracy of $0.596 \pm 0.109$, well above the $0.33$ chance level; at 21 exposures (7 per category) it climbed to $0.801 \pm 0.050$ , within 10\% of the maximum; and at 48 exposures (16 per category) it overlapped the peak within error ($0.866 \pm 0.037$ vs $0.886 \pm 0.033$).
In deployed settings, on-device few-shot recalibration can restore accuracy with minimal downtime and reference-gas consumption, lowering operating costs where data collection is expensive or logistically complex.

\subsection{Performance over time during an exposure}
\label{sec:rapid_inference}

% Systems operating in dynamic environments often require rapid decisions from short observation windows. 
% This is critical for safety monitors, where hazardous gases must be detected as early as possible~\cite{kim_development_2018},
% and for mobile platforms reacting to natural stimuli such as turbulent odour plumes, where low latency is essential for control and navigation~\cite{hassan_robotic_2024, jain_review_2025}. 

\begin{figure}
    \centering
    \includegraphics[width=0.5\linewidth]{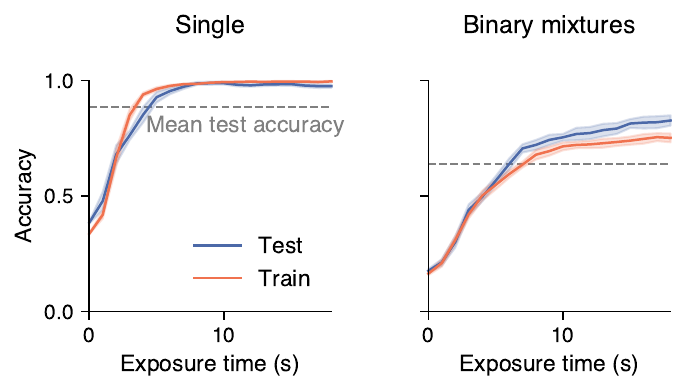}
    \caption{\textbf{Performance over exposure time.
    }
    Test accuracy (mean $\pm$ s.d. over 10 repetitions) versus time elapsed since the start of the exposure for the single-gas dataset (left) and the binary mixtures (right). The model is trained on the full training set and evaluated on individual test frames grouped by position. Dashed gray lines indicate the overall mean test accuracy. 
    }
    \label{fig:rapid_inference}
\end{figure}

Here we assessed how prediction accuracy depends on the time elapsed since gas presentation, with Fig.~\ref{fig:rapid_inference} reporting accuracy in both datasets as a function of exposure time. 
At $t=0$ accuracy was near chance ($0.385 \pm 0.005$ for single gases, $0.176 \pm 0.008$ for binary mixtures), as expected: the incoming gas takes a few seconds to displace the previous mixture from the sensor chamber and elicit a response, so the earliest readings still reflect the preceding stimulus.
Accuracy then rose rapidly, surpassing the mean test accuracy well before the sensors settled. 
While the time to a reliable prediction is set by the sensors' response time and the gas-delivery latency of our setup rather than the model, the network nonetheless predicts gas identity from the transient response rather than a stabilised fingerprint.

\subsection{Count sketch}
\label{sec:count_sketch}

In nature, tracking how often particular stimuli occur is a key computation that shapes behavioural decisions, helping organisms distinguish random occurrences from meaningful patterns in the environment~\cite{rankin_habituation_2009}. 
Familiarity with olfactory and visual scenes aids insects in staying on course while foraging and returning to the nest, whereas novel stimuli can trigger exploratory or escape responses that are critical for survival~\cite{collett_spatial_2013}. 
Analogously, robotic systems employ similarity search over visual scenes to support autonomous navigation in dense and cluttered environments~\cite{zhu_neuromorphic_2023}. 
Under the hood, these computations require storing and updating an internal look-up table of frequency estimates. 
In computer science, this problem is addressed by \textit{count sketches}~\cite{charikar_finding_2002, cormode_improved_2005}, hashing algorithms that extract approximate token frequencies from streamed data with compact memory and low computational cost.

\begin{figure}
    \centering
    \includegraphics[width=1.\linewidth]{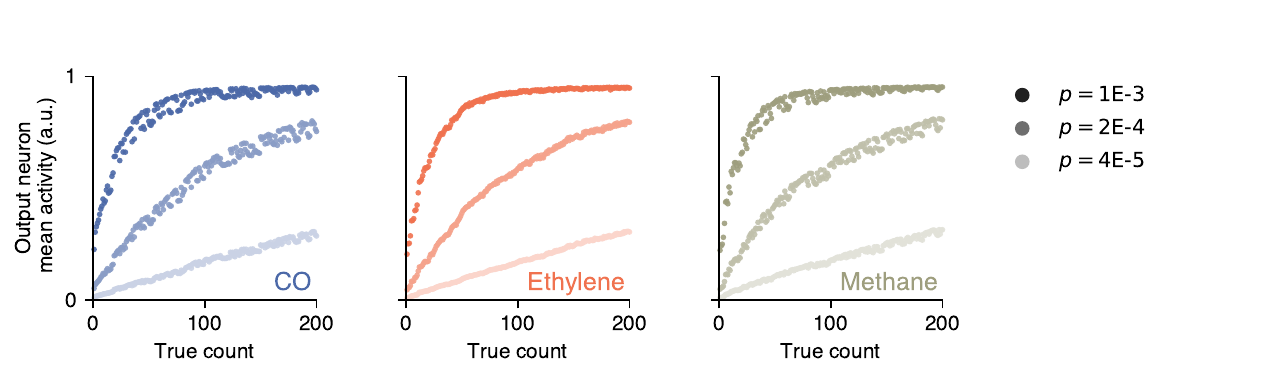}
    \caption{\textbf{Count sketch.
    }
    The mean activity of the read-out neuron for each mixture in a random sequence  of 200 presentations per gas of single gases. 
    }
    \label{fig:count_sketch}
\end{figure}

The expand-and-sparsify architecture naturally implements a count sketch through its sparse representations~\cite{dasgupta_neural_2018, dasgupta_neural_2022}.
Owing to the locality-preserving properties, similar sensor readings map to similar sparse vectors, which drives overlapping subsets of output synapses.
The read-out layer comprises a single neuron with always-on learning: any synapse whose presynaptic input is active is potentiated with probability $p$.
The potentiation probability controls the rate at which the output neuron's activity increases with repeated presentations of a given mixture, saturating once most synapses consistently recruited by a given mixture have been potentiated.
In this way, the frequency estimates are stored in the output weight matrix, approximating a class-identity count sketch with a distributed representation rather than an explicit table.
We demonstrate this counting capabilities in Fig.~\ref{fig:count_sketch} by tracking the read-out neuron activity during a sequence of single-gas presentations (CO, ethylene and methane). 
For each gas, the read-out activity increased linearly with the number of presentations until it saturated, effectively maintaining an online frequency estimate. 

A non-trivial feature of this mechanism is that it operates at the category level, keeping overall estimates per class rather than memorizing individual samples. 
This generalization is particularly relevant for continual learning systems, where this mechanism can be used to determine whether an input belongs to an already familiar category or represents a novel class that requires learning.

\subsection{On-chip hazardous emission detection}
\label{sec:on_chip_hed}

\begin{figure}
    \centering
    \includegraphics[width=1.\linewidth]{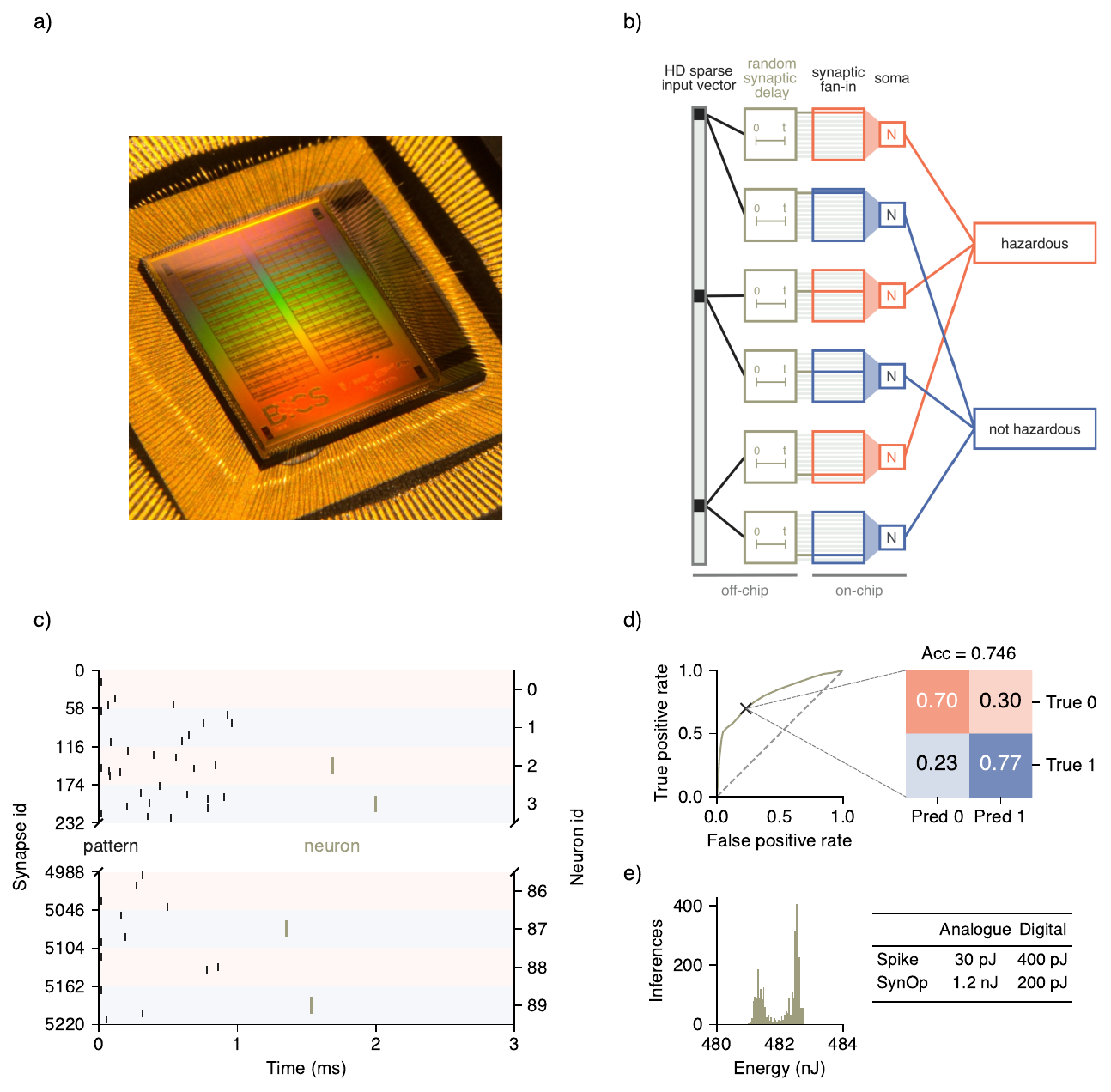}
    \vspace{5pt}
    \caption{\textbf{On-chip hazardous emission detection} 
    a) The custom mixed-signal neuromorphic processor TEXEL. It hosts 180 neurons distributed across two cores, with each neuron equipped with analogue synaptic circuits that emulate local plasticity mechanisms. 
    b) The network architecture mapped onto the chip. 
    \gls{hd} input vectors are distributed across neurons and their synaptic inputs, then converted into single spikes at nonzero vector entries with random delays relative to a teacher spike during training. 
    c) Raster plot of the input spike pattern for an inference sample, and the spiking responses of the read-out neurons.
    d) The \gls{roc} curve and the confusion matrix for the operating point with the highest Youden index (0: hazardous, 1: non-hazardous). 
    e) Histogram of the dynamic power consumption per inference across the test sequence. The table shows the energy cost per spike and Synaptic Operation (SynOp) in both analogue and digital domains, as reported in~\cite{greatorex_neuromorphic_2025}.
    }
    \label{fig:chip_mapping}
\end{figure}

We demonstrated the system on the mixed-signal neuromorphic processor TEXEL~\cite{greatorex_neuromorphic_2025} (Fig.~\ref{fig:chip_mapping}a).
We chose a \textit{hazardous emission detection} setting, a binary classification problem which requires to detect the presence of a hazardous gas from continuous sensor signals. 
Following the sampling protocol used in the previous sections, we presented random sequences of CO, ethylene and methane in 20 $\unit{\second}$ periods.
CO was designated the hazardous category (class 0) and the ethylene and methane the non-hazardous category (class 1).
The dataset comprised a training sequence of 10 exposures (2 for class 0, 8 for class 1; 190 sensor readings in total) and a test sequence of 150 exposures (50 for class 0, 100 for class 1; 2851 sensor readings in total).

Sensor data were encoded off-chip into sparse \gls{hd} vectors of length 2000 with the expand-and-sparsify scheme, at a fixed sparsity of 90$\%$.
The network ran on a single core of the chip, which provides 90 neurons with 58 synapses each, 54 of them plastic and binary-valued -- each synaptic weight can occupy only one of two states and changes state via on-chip \gls{stdp}.
Neuron, synapse and learning circuits all operate in the subthreshold analogue regime, giving extremely low per-event energy and a static (idle) power consumption of 27 $\unit{\micro\watt}$. 
Two readout populations of 40 neurons each were drawn at random from the good (non-spontaneously-active) neurons on the core, one for each class (Fig.~\ref{fig:chip_mapping}b).
Each HD vector was reshaped to a (40, 50) block and presented row by row, so that row $t$ drove plastic synapses on the $t$-th neuron of each readout population with a per-spike jitter of 0–1 $\unit{\milli\second}$. 
Before training, all plastic synapses were set to their low (zero) state. 
We trained on a class-balanced subset of 76 samples (38 per class) drawn from the training split, interleaved 0, 1, 0, 1, …, in a single pass. For each sample the input pattern was delivered to both populations, but only the target population additionally received a static teacher spike with a fixed delay, which reliably drove post-synaptic spiking activity. 
Pre-synaptic events on plastic synapses arriving immediately before this teacher-driven post-spike fell inside the \gls{stdp} potentiation window and switched those synapses from low to high; because the weights are binary, each plastic synapse can only be potentiated once and subsequent presentations of the same bit have no effect. The non-target population received the same vector shifted in time and with no teacher, so its plastic synapses was exposed to pre-spikes without a coincident post-spike and remained low.

During inference, learning was disabled and both readout populations received the same vector simultaneously (no time shift, no teacher) so that neither group could gain a transient advantage from accumulated global state. 
Fig.~\ref{fig:chip_mapping}c shows an example of pre- and post-synaptic spikes during inference.
Each pre-synaptic event arriving at a potentiated synapse contributed somatic current; output spikes from each population were summed over a 250 $\unit{\milli\second}$ window and the class was predicted from the comparison between the population activities
\begin{equation}
    y_i =
    \begin{cases}
        0, & \text{if } S_0 > S_1 - b_{thr} \\
        1, & \text{otherwise}
    \end{cases}
\end{equation}
where $y_i$ is the predicted label for sensor reading $i$, $S_0$ is the number of spikes of the neurons assigned to class 0 and $S_1$ is the number of spikes of the neurons assigned to class 1.
We tuned the classification threshold by adding a bias $b_{thr}$, yielding the \gls{roc} curve reported in Fig.~\ref{fig:chip_mapping}d.
We set the operating point maximizing the Youden's Index $J = \text{TPR} - \text{FPR}$, yielding an accuracy of 74.6$\%$.
Because both the encoding and the readout are sparse, the total spike count per inference is small, which combined with the subthreshold analog circuits results in an ultra-low-power implementation of the network at an average of 482$\unit{\nano\joule}$ per inference and a static power of 27$\unit{\micro\watt}$ (Fig.~\ref{fig:chip_mapping}e).

\section{Discussion}

\subsection{Ultra-low-power chemical sensing}

Edge monitoring devices -- portable multi-gas detectors for worker safety, residential alarms, indoor air quality systems, automotive cabin sensors and environmental monitoring devices -- must acquire, process and act on sensor data under tight energy and computing budgets.
The most energy (and cost) efficient devices 
% -- residential alarms and indoor air quality sensors -- 
typically employ mixed-signal \gls{asic} or 8-bit ultra-low-power \gls{mcu} 
for alarm triggering, baseline-drift compensation and end-of-life signalling, 
with power consumption in the range of a few hundred \unit{\micro\watt}~\cite{microchip_technology_8-bit_2009}.
Devices with
% Worker-safety devices have 
more complex requirements, such as periodic calibrations and device-to-device connectivity, typically employ 32-bit \glspl{mcu} with ARM Cortex-class processors or similar, with a power consumption of the order of a few \unit{\milli\watt}~\cite{nordic_semiconductors_32-bit_2016}.
Current solutions use dedicated electrochemical sensors for each target gas, and detection is based on comparison against fixed thresholds. Selectivity is built into the sensor design rather than in the processing.
% dedicated mixed-signal \glspl{asic}
% 8-bit ultra-low-power \glspl{mcu} 

However, many settings require selectivity to be built into the processing. 
\gls{mox} sensors are more cost-effective and have a longer lifetime than electrochemical sensors, and recent \gls{mems}-based fabrication has produced compact, low-cost units suited for large-scale deployment~\cite{asri_mems_2021}.  These are typically less selective, so replacing per-gas electrochemical sensors with other technologies shifts the selectivity from the sensor chemistry to the system level. 
Equally, complex odours are composed of a large number of gases and can hardly be identified by a fixed-threshold scheme, however selective the individual sensors. 
In both cases, the device must infer the category from the global sensor-array response.

Embedded machine learning has deployed classification models such as \gls{svm}~\cite{boubin_microcontroller_2019}, random forests~\cite{daghero_adaptive_2021} and neural networks~\cite{lai_cmsis-nn_2018} on ultra-low-power \glspl{mcu}.
The expand-and-sparsify architecture used in this work poses an alternative within a very tight power budget: the sparse \gls{hd} projection has been demonstrated on the RISC-V platform PULPv3 at $2 \unit{\milli\watt}$~\cite{montagna_pulp-hd_2018}, and we demonstrated the read-out layer on the mixed-signal chip TEXEL at $482 \unit{\nano\joule}$ per inference ($27\unit{\micro\watt}$ static-dominated total at a sampling rate of $1\unit{\hertz}$). 
End-to-end, the system runs at the order of $2 \unit{\milli\watt}$, an order of magnitude below a comparable pipeline on an ARM Cortex M4 at $20\unit{\milli\watt}$~\cite{montagna_pulp-hd_2018}.
This gap reflects the move from a general-purpose processor to specialized low-power hardware, a move made possible by the architectural features -- sparse binary projections and Hebbian learning --  that fit naturally onto such substrates, where dense and high-precision floating point inference is too costly.

\subsection{Toward field-deployable gas classification}

Common gas classification models are typically developed under a predefined experimental protocol~\cite{zhang_new_2022, peng_gas_2018, wei_development_2019, wang_exploiting_2023, du_vision_2024, kwon_efficient_2021, kang_high_2022, xiong_gas_2021, vanarse_application_2022, huo_bio-inspired_2023, xue_novel_2024}:
the sensor array is exposed to a clean-air baseline before and after each sample, and the trace is segmented into well-defined baseline, rise, steady-state and recovery phases.
This protocol captures the signal transients together with the steady state response, but the issues arising from this approach make it a poor match to on-field applications.
Because a prediction becomes available only after a complete baseline–exposure–recovery cycle, the scheme cannot meet the low-latency requirements of applications such as hazardous-gas monitoring or mobile platforms.
It also assumes stationary statistics within each exposure, whereas natural settings present turbulent plumes with complex spatial and temporal structure~\cite{connor_quantification_2018}, so odours from the same source can differ substantially with airflow and local concentration gradients. Models validated only under baseline-exposure-recovery cycles do not necessarily transfer to such conditions.
Online operation is further limited by pre-processing that relies on pre-learnt statistics, which may differ from those encountered during deployment, and is often computed offline.

Previous work has focused on field operation, continuously estimating gas identity or concentration under fast-varying inputs, using linear and non-linear classifiers~\cite{pardo_time_2000, de_vito_gas_2007, trincavelli_odour_2009}, reservoir computing~\cite{fonollosa_reservoir_2015} and bio-inspired spiking networks~\cite{diamond_classifying_2016}.
Our system operates in the same regime: it continuously estimates gas identity from the transient response, and its sampling protocol relaxes the stationarity assumption -- mixtures are introduced in a random succession, so the initial condition of each exposure is set by the preceding stimulus, yielding a broad distribution of trajectories per category rather than steady-state patterns.
It also operates directly on the raw sensor readings, removing the preprocessing mismatch between laboratory and deployment conditions.
The key differentiator of our system is that it does so while learning online and on-chip, with a lightweight network suited to ultra-low-power neuromorphic hardware---an important step towards on-field, online gas classification at the edge.

\subsection{Few-shot calibration}

Gas sensors degrade over their operating lifetime through baseline drift, sensitivity loss and selectivity changes~\cite{vergara_chemical_2012, dennler_drift_2022}, therefore deployed devices require periodic recalibration. 
Because acquiring labelled reference data is expensive, learning the category distribution from only a few labelled samples -- an active and challenging problem in machine learning~\cite{parnami_learning_2022, song_comprehensive_2023} -- is particularly relevant in this setting. 
Many existing embedded machine learning approaches deploy offline-trained, frozen models~\cite{boubin_microcontroller_2019, daghero_adaptive_2021, lai_cmsis-nn_2018, gkogkidis_tinyml-based_2022, krayden_cmos-mems_2022}: recalibration then requires collecting fresh labelled data from each unit, retraining off-device, and redeploying the updated model. 
In contrast, the on-chip learning in the expand-and-sparsify network enables on-device recalibration, where a controlled re-exposure to reference gases is sufficient to update the category prototypes. Its data efficiency allows recalibration with a very small number of exposures, minimizing device downtime and calibration-gas use.

\section{Methods}
\label{sec:methods}

\subsection{Experimental setup}
\label{sec:experimental_setup}

\subsubsection{Sensor array}
\begin{figure}
    \centering
    \includegraphics[width=1.\linewidth]{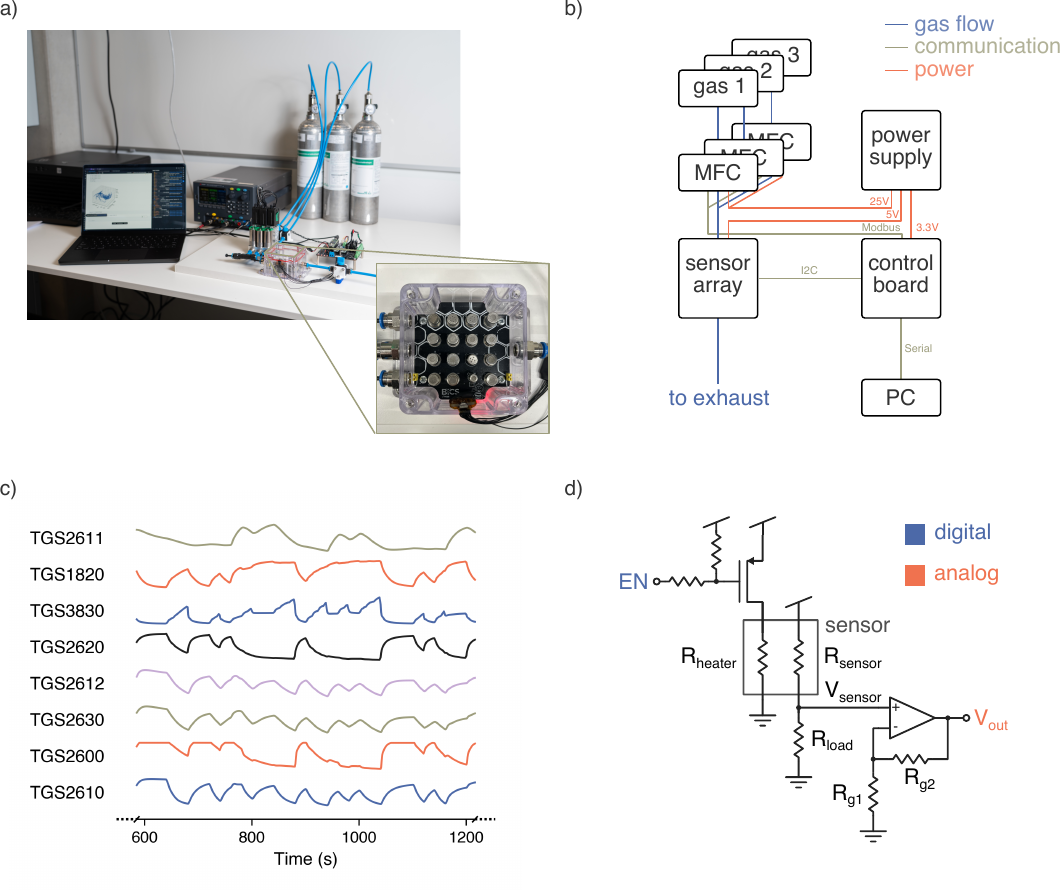}
    \vspace{5pt}
    \caption{\textbf{Experimental setup.
    }
    a) Photography taken during a recording session.
    b) Schematic view of the experimental setup.
    c) Sensor array signals. Traces recorded from the 8 active sensors during 600s of continuous sampling.
    d) Sensor read-out circuit. The sensor acts as a variable resistor. The transistor, controlled by the digital enable signal (EN), switches the heater on/off. The sensor and the load resistor form a voltage divider. The voltage $\text{V}_\text{sensor}$ is fed into a non-inverting amplifier with a gain set by $\text{R}_\text{g1}$ and $\text{R}_\text{g2}$.
    }
    \label{fig:methods}
\end{figure}

The sensor array consists of 16 commercial \gls{mox} sensors (see Table~\ref{tab:sensor_array}), integrated on a custom \gls{pcb} together with the control and read-out circuits (Fig.~\ref{fig:methods}).
The heater in each sensor can be switched on and off via the transistor M1, driven by digital \gls{pwm} signals.
The sensor output is measured using a voltage divider formed by the load resistor $R_\text{load}$ and the intrinsic sensor resistance $R_\text{sensor}$, such that the node voltage $V_{sensor}$ is given by

\begin{equation}
    V_\text{sensor} = \frac{R_\text{load}}{R_\text{load} + R_\text{sensor}}V_\text{ref}.
\end{equation}

The value of $R_\text{load}$ is chosen according to the nominal operating resistance of the sensor, following the manufacturer’s specifications. 
Subsequently, the output is amplified using an \gls{opamp} with gain set by resistors $R_{g1}$ and $R_{g2}$:

\begin{equation}
    V_\text{out} = \frac{R_{g1}+R_{g2}}{R_{g1}} V_\text{sensor} = \frac{R_{g1}+R_{g2}}{R_{g1}}\frac{R_\text{load}}{R_\text{load} + R_\text{sensor}}V_\text{ref}
\end{equation}
so that $V_\text{out}$ fits within the input range of the \glspl{adc}. 
% Values of $R_\text{load}$, $R_{g1}$ and $R_{g2}$ for each sensor are listed in (ref).

Signal acquisition is performed using four 4-channel \glspl{adc} (ADS114, Texas Instruments), which communicate with the control unit via the \gls{i2c} bus. 
Digital PWM signals for heater control are provided by an I/O expander (PCF8575, Texas Instruments), also addressed through \gls{i2c}.

\subsubsection{Gas delivery system}
The test gases are supplied in 110~L pressurized cylinders (Air Products NL). Each cylinder is equipped with a fixed-flow regulator set to $1.5 \unit{\liter\per\minute}$. 

The system comprises three input lines, each regulated by a \gls{mfc} (Bronkhorst FF-A005). The three input lines converge in a single line from which the gas flow is directed to the sensor array. The sensor array is placed in a hermetic container of $\sim0.05\unit{\liter}$. The small internal volume and open geometry of the sensor chamber promotes rapid gas diffusion across the sensors. Finally, an exhaust line vents the gas safely to a fume extractor for safe disposal.

\subsubsection{Control unit}
The control unit consists of a custom \gls{pcb} with a Teensy~4.1 \gls{mcu}. It handles sensor reading and configuration of the \glspl{adc} and I/O expander via \gls{i2c}, and controls the \glspl{mfc} over \gls{uart} through a UART-to-ModBus converter.

The board integrates power management circuits with AC–DC converters and voltage regulators that supply $24\unit{\volt}$, $5\unit{\volt}$, $3.3\unit{\volt}$, and $2.3\unit{\volt}$. It can operate directly from $240\unit{\volt}$ AC or from external $24\unit{\volt}$ and $5\unit{\volt}$ DC inputs, with the on-board regulators generating the lower-voltage rails.

\subsection{Data acquisition}
\label{sec:sampling_procedure}
We generate random sequences with a predefined number of samples. The mixtures used are ethylene, carbon monoxide, methane and their binary mixtures, at 2000 ppm with air as the solvent gas. 
This gas selection is widely used in odour classification literature~\cite{vergara_performance_2013,fonollosa_chemical_2014, fonollosa_reservoir_2015, peng_gas_2018, wei_development_2019, xiong_gas_2021, lee_classifying_2022, mahdavi_2d-cnn_2022, pan_lightweight_2022, zhong_research_2022,}. 
Characteristic time constants in the sensor responses lie in the $20$ -- $40\unit{\second}$ range (Fig.~\ref{fig:extended_sensor_characterization}).
Therefore the mixtures are continuously delivered to the sensor array in $20\unit{\second}$ intervals, guaranteeing steady-states are rarely reached.
Single-gas mixtures are sampled by setting the corresponding \gls{mfc} at $1.5\unit{\liter\per\minute}$, while binary mixtures are sampled by setting two lines at $0.75\unit{\liter\per\minute}$ each, resulting in a 1:1 mixture in volume. 

The heater operating voltage is maintained at $5\unit{\volt}$, and the \glspl{adc} are sampled at $1\unit{\hertz}$ via the \gls{i2c} interface. 
Only a subset of sensors (3, 4, 7, 8, 9, 11, 12, 13) is used in this work, corresponding to the eight sensors with the strongest responses to the test gases. Fig.~\ref{fig:methods}c shows the output traces for $600\unit{\second}$ of continuous sampling.

\subsection{The expand-and-sparsify network}
\label{sec:expand_and_sparsify}

% [The e&s architecture]
At a given $t$, the sensor array outputs are low-dimensional vectors $\mathbf{x}(t) \in \mathbb{R}^{N}$. These are projected into a high-dimensional space with a random binary matrix $\bold{W}~\in~\{0,1\}^{N \times M}$, so
\begin{equation}
    \mathbf{z}_\text{dense}(t) = \bold{W}^{T} \mathbf{x}(t), \quad \mathbf{z} \in \mathbb{R}^{M}.
\end{equation}
The coefficients of $\textbf{W}$ are given by $\text{W}_{ij} \equiv \text{Bernoulli}(p_{HD})$. Therefore, the mean sparsity of the projection matrix ${\langle s_\textbf{W}\rangle=1-p_{HD}}$ is defined by $p_{HD}$.

\begin{figure}
    \centering
    \includegraphics[width=0.5\linewidth]{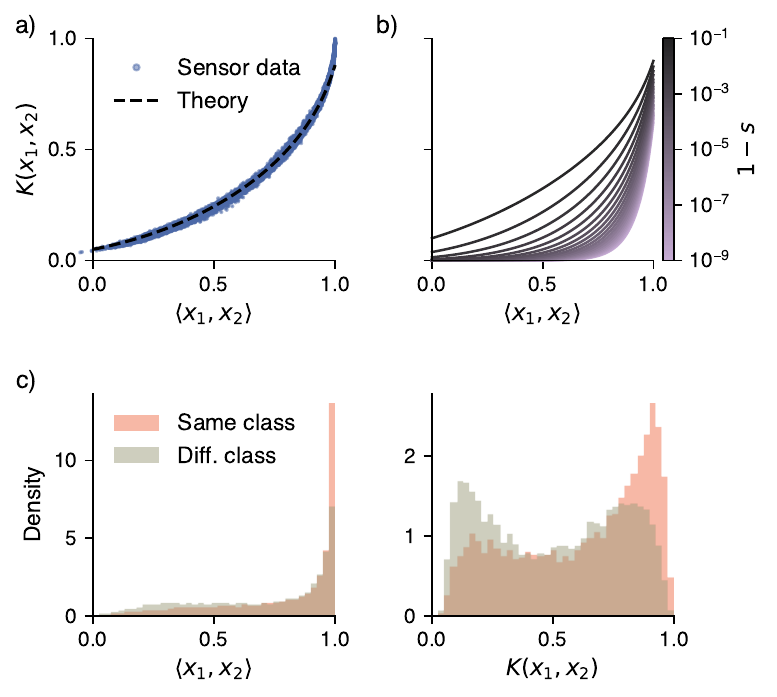}
    % \vspace{5pt}
    \caption{\textbf{The kernel induced by the expand-and-sparsify architecture.} a) The measured kernel for random pairs of expanded sensor read-outs ($n_{pairs}=10000$, $d_{input}=136$, $p_{HD} = 0.2$ and $1-s = 0.2$) compared against the theoretical kernel. b) The theoretical kernel as a function of the sparsity of the final representation, determined by $k$ in the $k$-WTA.
    c) Histograms representing the inner products between random pairs of sensor read-outs (left) and its sparsified representations (right).
    }
    \label{fig:kernel}
\end{figure}

Following the high-dimensional projection, $z_\text{dense}$ is sparsified with a \gls{kwta} mechanism
\[
\mathbf{z}_\text{sparse} = \mathcal{H}_k(\mathbf{z}_\text{dense}) \quad \text{(top-$k$)}.
\]
The $k$ largest elements of $\mathbf{z}_\text{dense}$ are set to 1, and the remainder to 0. This results in a binary vector
\begin{equation}
    \mathbf{z}_\text{sparse}(t) \in \{0,1\}^M \quad\text{with sparsity}\quad s = 1 - \frac{k}{M},
\end{equation}
which serves as the sparse \gls{hd} representation of the sensor output.
The \gls{kwta} mechanism approximates a kernel function $K$~\cite{hugh_thesis}:
\begin{equation}
    \mathbb{E}[ \mathbf{z}_\text{sparse}(x_1) \cdot \mathbf{z}_\text{sparse}(x_2)] = M(1-s)K(\mathbf{x}_{1}, \mathbf{x}_{2}), 
\end{equation}
where the kernel is given by (Fig.~\ref{fig:kernel}a)
\begin{equation}\label{eq: kernel}
    K(\mathbf{x}_{1}, \mathbf{x}_{2}) = (1-s) + \sum_{n=1}^{\infty} \frac{r^n}{n!}\left[\frac{\phi(t)\He_{n-1}(\theta)}{(1-s)}\right]^2 .
\end{equation}
The coefficient $r=\langle \vec{x}_1, \vec{x}_2 \rangle$ is the input correlation, $\phi(x)~=~\frac{1}{\sqrt{2\pi}}e^{-\frac{x^2}{2}}$ is the standard normal probability density function, $\He_{n}$ is the $n$-th Hermite polynomial, and $\theta = \Phi^{-1}(s)$ where $\Phi$ is the \gls{cdf} of the standard normal distribution. 

The kernel shape is controlled by the sparsity of the final representation, set by $k$ in the $k$-WTA (Fig.~\ref{fig:kernel}b).
Consistent with the behaviour of non-linear \gls{svm} kernels, our projected samples exhibit increased inter-class decorrelation and tighter intra-class aggregation (Fig.~\ref{fig:kernel}c).

\subsection{Linear read-out}
\label{sec:read_out}

\subsubsection{Binary weights trained with Hebbian learning}

The sparse representation generated by the expand-and-sparsify architecture is projected to the read-out layer via a binary matrix $\bold{W}_\text{out}~\in~\{0,1\}^{M \times L}$, so
\begin{equation}\label{eq:out}
    \mathbf{o}(t) = \bold{W}_\text{out}^{T} \mathbf{z}_\text{sparse}(t), \quad \mathbf{o} \in \mathbb{R}^{L}
\end{equation}
where $M$ is the dimensionality of the sparse representation and $L$ the dimensionality of the output layer.
This implements a linear separation of the sparse representations $\textbf{z}_\text{sparse}$, with category estimations given by Eq.~\ref{eq:out}.
The columns of $\bold{W}_\text{out}$ define the orientations of the hyperplanes separating the regions associated with each category.

There is a binary teacher signal $\mathbf{l}(t) \in \{0,1\}^L$ that modulates the learning of the output synapses. $\bold{W}_\text{out}$ is initialized with all coefficients set at 0, and the weight potentiations are given by
\begin{equation}
    \Delta \mathbf{W}_\text{pot} = \mathcal{B} \cdot (\textbf{z}_\text{sparse} \otimes \textbf{l}) \circ (\mathbf{1} - \mathbf{W}_\text{out}),
\end{equation}
where ${\mathcal{B} \sim \text{Bernoulli}^{(M \times L)}(p) \in \{0, 1\}^{M \times L} }$, ${\mathbf{1} \in \{1\}^{M\times L}}$ is a matrix of ones, $\otimes$ is the outer product and $\circ$ the Hadamard product. Synapses have a probability $p$ of potentiating if the presynaptic unit is active ($z_{i} = 1$) and the learning signal to the post-synaptic unit is active ($l_{j} = 1$). 

This learning rule implements Hebbian synaptic plasticity in the output synapses~\cite{hebb_organization_1945, sejnowski_book_nodate}. The output layer accumulates evidence to estimate each class distribution in the sparse \gls{hd} representation. The non-deterministic weight updates allow to do so by requiring on average multiple activations of the presynaptic unit to flip the synapse. This saves storing traces of past activity per synapse.

\subsubsection{High-resolution weights trained with gradient-descent}
To establish an upper bound on the linear read-out performance, we trained a linear \gls{svm} read-out with high-resolution weights (64-bits) using gradient-descent.
The category estimates are given by
\begin{equation}\label{eq:cat_estimation}
    \mathbf{o}(t) = \mathbf{W}_\text{out}^{T} \mathbf{z}_\text{sparse}(t) + \mathbf{b}, \quad \mathbf{o} \in \mathbb{R}^{L}
\end{equation}
where $\mathbf{b}$ is a constant bias to each read-out neuron. 
The weight matrix $\mathbf{W}_\text{out}$ and the biases $\mathbf{b}$ are optimized using a Crammer-Singer multi-class hinge loss with L2 regularization
\begin{equation}\label{eq:cs-loss}
\begin{aligned}
\mathcal{L}(\mathbf{W}_\text{out}, \mathbf{b}) &= \frac{1}{n}\sum_{i=1}^{n}\sum_{\substack{j=1 \\ j \neq y_i}}^{K} \max\!\bigl(0,\; \text{o}_{i,j} - \text{o}_{i,y_i} + 1\bigr) +\\
&\phantom{=}+ \frac{1}{2\,C\,n}\,\lVert \mathbf{W}_\text{out} \rVert_F^{2},
\end{aligned}
\end{equation}
where $n$ is the number of data points, $K$ the number of categories, $\text{o}_{i,j} \equiv \text{o}(t=i)_j$ is the $j$-th component of the read-out activity at step $i$, and $y_i$ is the corresponding label, $C$ is the regularization parameter and $\lVert \cdot\rVert_{F}$ denotes the Frobenius norm. 
The read-out layer is trained for 200 epochs using the Adam optimizer.

\subsection{Network parametrisation}
\label{sec:parametrisation}

All software experiments share one expand-and-sparsify configuration and differ only in the read-out and its potentiation probability, as summarised in Table~\ref{tab:parametrisation}.
The projection operates on the $N = 136$ sensor expansion (Section~\ref{sec:dimensionality_expansion}, Table~\ref{tab:filtered_expansion}). Its two free parameters, the matrix density $p_{HD}$ and the $k$-\gls{wta} sparsity $s$, were fixed by a grid search on the classification task ($p_{HD} \in [0.025, 0.975]$, $k/M \in [0.025, 0.5]$, potentiation $p \in \{0.01, 0.1\}$, 10 seeds per point) and then reused unchanged for the few-shot, exposure-time and count-sketch experiments;
the repeated rows in Table~\ref{tab:parametrisation} reflect this reuse rather than independent tuning. 
Each configuration was evaluated over 10 repetitions with independent random projections.

The gradient-descent read-out keeps the shared projection and replaces the binary Hebbian read-out with the 64-bit linear classifier trained by Adam. 
The count-sketch read-out is a single always-on Hebbian neuron with always-on learning, and its potentiation probability is set one to three orders of magnitude below the classification value so that the weights accumulate slowly and the neuron integrates presentation frequency rather than classifying. 
In the on-chip experiment its shorter representation ($M = 2000$) is set by the chip constraints, whereas its sparsity was selected, as for the other experiments, by a grid search over the active fraction $k/M$ and the number of training presentations.

The linear and non-linear SVM baselines were tuned independently. Both were fit and evaluated on the same sequential train/test split used for the network--600 presentations for training and 300 for testing on each dataset-- so that all methods are compared under identical conditions; because each fit is deterministic given its training set, the \gls{svm} entries in Table~\ref{tab:classification} are reported from a single fit (std $=0$). The linear \gls{svm} used the Crammer--Singer multi-class hinge loss of Eq.~\ref{eq:cs-loss}, minimised by gradient descent (Adam, 200 epochs), with its regularisation parameter $C$ selected by grid search over $C \in [10^{-2}, 10^{10}]$ (ten logarithmically spaced values), yielding $C \approx 10^{-2}$ for single gases and $C \approx 0.22$ for binary mixtures. The \gls{rbf} \gls{svm} hyperparameters were selected by grid search over the same range of $C$ and over $\gamma \in [10^{-18}, 10^{-3}]$ (sixteen logarithmically spaced values), the grid extended towards small $\gamma$ after the optimum reached the lower edge of the original range. Each kernel was optimised separately on each dataset's own split, yielding $C \approx 2.2 \times 10^{7}$, $\gamma = 10^{-12}$ for single gases and $C \approx 2.2 \times 10^{7}$, $\gamma = 10^{-11}$ for binary mixtures.

\subsection{Input dimensionality effect on the kernel mismatch}
\label{sec:dimensionality_expansion}

\begin{figure}
    \centering
    \includegraphics[width=\linewidth]{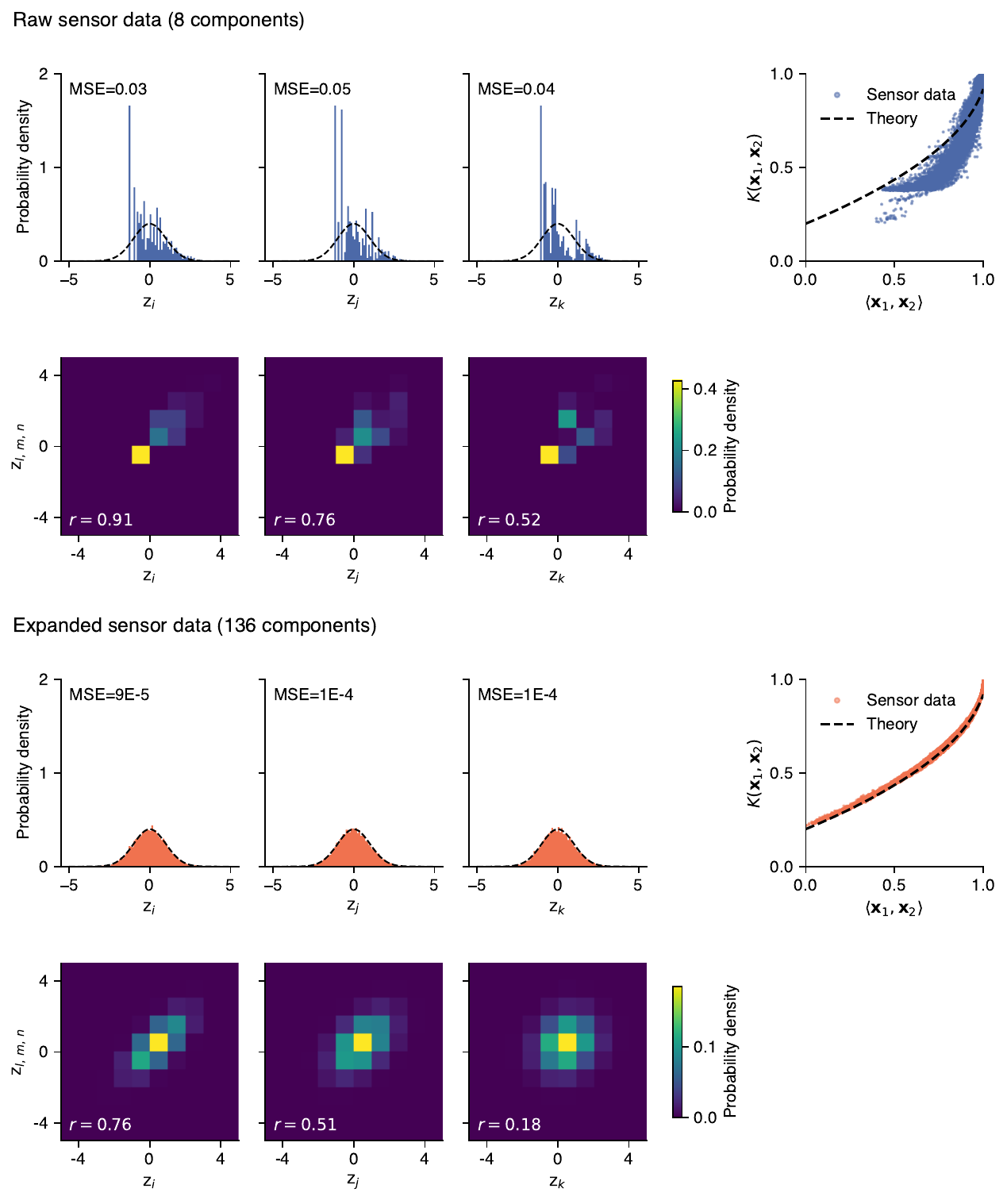}
    % \vspace{5pt}
    \caption{\textbf{Effect of the input dimensionality on the projection statistics.
    }
    Top: raw 8-channel sensor traces. Bottom: 136-dimensional expanded traces. Histograms show the empirical density of the $\mathbf{z}_{\text{dense}}$ components for 3 random samples, with the MSE to the standard normal annotated ($n=10000$). Heat-maps show the joint density for 3 random pairs of samples ($n=10000$), with the Pearson correlation between the samples annotated. Scatter plots show the measured kernel versus the theoretical one described by Eq.~\ref{eq: kernel} for random pairs of samples ($n=10000$).
    }
    \label{fig:kernel_dim_expansion}
    
\end{figure}

The kernel derivation in Eq.~\ref{eq: kernel} assumes that the components of the \gls{hd} projection $\mathbf{z}_\text{dense}$ are normally distributed:
each component $\text{z}_{i} = \sum_{j}\text{W}_{ji}\text{x}_j(t)$ is the sum of a random subset of the input components $\text{x}_j(t)$, selected by the nonzero entries of the projection weights $\text{W}_{:,i}$. 
The number of pooled terms is $\sim p_\text{HD} N$, where $p_\text{HD}$ is the density of \text{W} and $N$ the input dimensionality.
By the \gls{clt}, the $\text{z}_{i}$ distribution tends to a normal as the number of pooled terms grows.
For low-dimensional inputs the \gls{clt} assumption is poorly met and the realised kernel deviates from Eq.~\ref{eq: kernel}. 
Fig.~\ref{fig:kernel_dim_expansion} illustrates this effect. We estimate univariate and joint densities of $\mathbf{z}_\text{dense}$ components from $n=10000$ random pools of input channels given by $\text{Bernoulli}(p_{HD})$. 
With the eight sensor traces, the $\text{z}_{i}$ distributions deviate strongly from a normal ($\text{MSE} \sim 10^{-2}$), while expanding the input dimensionality brings the $\text{z}_{i}$ distributions close to normal ($\text{MSE} \sim 10^{-4}$). 
The input dimensionality determines the mismatch between the inner products of the transformed samples and the theoretical kernel given by Eq.~\ref{eq: kernel}

We next assess how this mismatch affects the linear separability of the transformed sensor data, distinguishing between two contributions: dimensionality and information content.
Tab.~\ref{tab:filtered_expansion} reports linear classification accuracy on the dense sensor data $\mathbf{x}_\text{dense}$ and on the transformed $\mathbf{z}_\text{sparse}$ for a range of sensor output dimensionality expansions.
Linear separation of $\mathbf{x}_\text{dense}$ shows that only adding temporal derivatives improves the separation accuracy: 
the expansions adding pairwise differences or ratios yield unchanged or slightly lower accuracy compared to expansions with the same number of temporal derivatives but less pairwise differences or ratios.
The backward differences supply information about the recent history of the sensor response -- which is otherwise unavailable to an instantaneous read-out -- while the pairwise differences and ratios increase the data redundancy and can potentially amplify noise without adding any new information.
In contrast, both types of sensor dimensionality expansion improve the linear separability of the transformed $\mathbf{x}_\text{dense}$. Despite carrying no additional information, increasing the number of components pooled by each projection reduces the mismatch with the theoretical kernel.
When the mismatch is large a given input pair no longer maps to a single inner product in the transformed space but to a broad distribution around the theoretical value. This non-deterministic mapping distorts the local geometry of the input space, complicating the class boundaries and degrading linear separability.
This result highlights a counter-intuitive effect: adding redundant and noisy terms improves the data distribution in the \gls{hd} feature space through their effect on the projection statistics alone.

\FloatBarrier

\section*{References}
\addcontentsline{toc}{section}{References}

% \balance
\printbibliography[heading=none]

\section*{Acknowledgements}

\addcontentsline{toc}{section}{Acknowledgements}

The authors would like to thank Vincent Jassies and Herman Adema for their technical support in developing PCBs and firmware for the gas sensing setup.
Additionally we would like to thank Madison Cotteret for his help soldering the PCBs.
The authors would like to acknowledge the financial support of the CogniGron research center and the Ubbo Emmius Funds (Univ. of Groningen).

\section*{Author contributions}

Conceptualisation - T. J. P., H. G., E. C.; 
Methodology - T. J. P., H. G.;
Software/Hardware - T. J. P., H. G.; 
Investigation - T. J. P.; 
Writing - T. J. P., H. G., E. C.; 
Visualisation - T. J. P., H. G., E. C.;  
Supervision - E. C.

\section*{Competing interests}
The authors declare no competing interests.

\clearpage
\section*{Tables}
\addcontentsline{toc}{section}{Tables}

\begin{table}[h]
\centering
\resizebox{\textwidth}{!}{%
\begin{tabular}{@{}ll cccc@{}}
\toprule
         &                 & \multicolumn{2}{c}{Train accuracy}
                           & \multicolumn{2}{c}{Test accuracy} \\
         \cmidrule(lr){3-4} \cmidrule(l){5-6}
Method   & Readout         & Single ($n{=}600$)
                           & Binary ($n{=}600$)
                           & Single ($n{=}300$)
                           & Binary ($n{=}300$) \\
\midrule
Expand \& sparsify        & Hebbian -- binary weights      & $0.859 \pm 0.009$ & \textbf{\boldmath $0.602 \pm 0.014$} & $0.881 \pm 0.014$ & \textbf{\boldmath $0.645 \pm 0.017$} \\
\midrule
Linear SVM                & NA                             & \textbf{\boldmath$0.902 \pm 0.000$} & $0.551 \pm 0.000$ & \textbf{\boldmath $0.923 \pm 0.000$} & $0.563 \pm 0.000$ \\
\midrule
Expand \& sparsify        & Linear classifier              & $0.929 \pm 0.004$ & $0.849 \pm 0.004$ & $0.952 \pm 0.003$ & $0.791 \pm 0.008$ \\
\midrule
RBF SVM                   & NA                             & $0.927 \pm 0.000$ & $0.853 \pm 0.000$ & $0.940 \pm 0.000$ & $0.800 \pm 0.000$ \\
\bottomrule
\end{tabular}%
}
\vspace{5pt}
\caption{Train and test accuracy for the expand-and-sparsify architecture
compared to linear and non-linear SVM methods. Network read-outs are reported as
mean $\pm$ std over 10 repetitions with independent random projections; the
linear and \gls{rbf} \gls{svm} baselines are deterministic fits on the same sequential
train/test split as the network and are therefore reported from a single run
(std $= 0$).}
\label{tab:classification}
\end{table}

\begin{table}[h]
\centering
\resizebox{0.845\linewidth}{!}{%
\begin{tabular}{lccccccc}
\toprule
Experiment & $M$ & $p_{HD}$ & $k$ & $s$ & Read-out & $L$ & $p$ \\
\midrule
Classification (Hebbian) & $10\,000$ & $0.025$ & $5000$ & $0.5$ & Hebbian, 1-bit & 3\,/\,6 & $0.01$ \\
Classification (GD)      & $10\,000$ & $0.025$ & $5000$ & $0.5$ & GD, 64-bit     & 3\,/\,6 & -- \\
Few-shot                 & $10\,000$ & $0.025$ & $5000$ & $0.5$ & Hebbian, 1-bit & 3\,/\,6 & $0.01,\,0.1$ \\
Exposure time            & $10\,000$ & $0.025$ & $5000$ & $0.5$ & Hebbian, 1-bit & 3\,/\,6 & $0.01$ \\
Count sketch             & $10\,000$ & $0.025$ & $5000$ & $0.5$ & Hebbian, 1-bit & 1       & $10^{-3}$--$4\!\times\!10^{-5}$ \\
On-chip (TEXEL)          & $2000$    & $0.025$ & $200$  & $0.9$ & \gls{stdp}, binary & 2   & -- \\
\bottomrule
\end{tabular}%
}
\caption{Network parametrisation per experiment. 
$M$: dimensionality of the HD projection; 
$p_\mathrm{HD}$: density of the random binary projection matrix; 
$k$: number of $k$-WTA winners; 
$s = 1 - k/M$: sparsity of the HD representation; 
$L$: read-out dimensionality (single gases / binary mixtures); $p$: potentiation probability of the read-out weights. 
All software experiments used the $N = 136$ input expansion.}
\label{tab:parametrisation}
\end{table}

\begin{table}[h]
\centering
\resizebox{0.89\linewidth}{!}{%
\begin{tabular}{lccccc}
\toprule
 & & \multicolumn{2}{c}{Test accuracy ($x_{\mathrm{dense}}$)} & \multicolumn{2}{c}{Test accuracy ($z_{\mathrm{sparse}}$)} \\
\cmidrule(lr){3-4} \cmidrule(lr){5-6}
Expansion & $n_{\mathrm{dense}}$ & Single gases & Binary mixtures & Single gases & Binary mixtures \\
\midrule
raw & 8 & $0.793 \pm 0.083$ & $0.390 \pm 0.054$ & $0.840 \pm 0.006$ & $0.533 \pm 0.015$ \\
\midrule
$\partial ^1$ & 16 & $0.919 \pm 0.069$ & $0.598 \pm 0.074$ & $0.969 \pm 0.001$ & $0.704 \pm 0.008$ \\
$8\text{D}$ & 16 & $0.795 \pm 0.070$ & $0.368 \pm 0.057$ & $0.888 \pm 0.014$ & $0.613 \pm 0.015$ \\
$8\text{R}$ & 16 & $0.807 \pm 0.068$ & $0.395 \pm 0.061$ & $0.837 \pm 0.011$ & $0.537 \pm 0.006$ \\
\midrule
$\partial ^1+\partial ^2$ & 24 & $0.933 \pm 0.057$ & $0.622 \pm 0.074$ & $0.969 \pm 0.002$ & $0.721 \pm 0.015$ \\
$\partial ^1 + 8\text{D}$ & 24 & $0.904 \pm 0.079$ & $0.556 \pm 0.077$ & $0.968 \pm 0.003$ & $0.756 \pm 0.027$ \\
$\partial ^1 + 8\text{R}$ & 24 & $0.912 \pm 0.075$ & $0.605 \pm 0.073$ & $0.971 \pm 0.001$ & $0.715 \pm 0.014$ \\
$8\text{D} + 8\text{R}$ & 24 & $0.798 \pm 0.073$ & $0.376 \pm 0.061$ & $0.883 \pm 0.018$ & $0.609 \pm 0.015$ \\
\midrule
$\partial ^1+\partial ^2 + 8\text{R} + 8\text{D}$ & 40 & $0.917 \pm 0.067$ & $0.577 \pm 0.075$ & $0.971 \pm 0.002$ & $0.783 \pm 0.012$ \\
\midrule
$\partial ^1+\partial ^2 + 56\text{R} + 56\text{D}$ & 136 & $0.888 \pm 0.087$ & $0.501 \pm 0.083$ & $0.965 \pm 0.008$ & $0.789 \pm 0.010$ \\
\bottomrule
\end{tabular}%
}
\vspace{5pt}
\caption{
Linear SVM accuracy on $\mathbf{x}_{\text{dense}}$ and $\mathbf{z}_{\text{sparse}}$ for different expansions of the sensor traces. $\partial ^1$ and $\partial ^2$ denote the first- and second-order temporal derivatives approximated with backward differences. $n\text{D}$ and $n\text{R}$ denote $n$ pairwise differences and ratios between sensor traces respectively. $n_\text{dense}$ is the resulting dimensionality. Test accuracy is reported as mean $\pm$ standard deviation over 100 repetitions.
}
\label{tab:filtered_expansion}
\end{table}

\clearpage
\section*{Figures}
\addcontentsline{toc}{section}{Figures}

% \begin{figure}
%     \centering
%     \includegraphics[width=0.5\linewidth]{figs/sampling_trajectories.pdf}
%     \vspace{5pt}
%     \caption{\textbf{Trajectories in feature space.}
%     Representation of the sample trajectories in feature space during exposure-recovery cycles and in continuous sampling mode.
%     }
%     \label{fig:trajectories}
% \end{figure}

% \begin{figure}
%     \centering
%     \includegraphics[width=0.5\linewidth]{figs/readout_circuit_nolegend.pdf}
%     \vspace{5pt}
%     \caption{\textbf{Sensor read-out circuit.}
%     The sensor acts as a variable resistor. The transistor, controlled by the digital enable signal (EN), switches the heater on/off. The sensor and the load resistor form a voltage divider. The voltage $\text{V}_\text{sensor}$ is fed into a non-inverting amplifier with a gain is set by $\text{R}_\text{g1}$ and $\text{R}_\text{g2}$.
%     }
%     \label{fig:readout_circuit}
% \end{figure}

% \begin{figure}
%     \centering
%     \includegraphics[width=0.5\linewidth]{figs/sensor_signal.pdf}
%     \caption{\textbf{Sensor array signals.}
%     Traces recorded from the 8 active sensors during 400s of continuous sampling.
%     }
%     \label{fig:circuits_sensor_signal}
% \end{figure}

\onecolumn
\FloatBarrier
\newpage

\begin{center}
\textsc{Supplementary Material}
\end{center}

\renewcommand\thefigure{S\arabic{figure}}
\renewcommand\thesection{S\arabic{section}}
\renewcommand{\theHfigure}{S\arabic{figure}}
\renewcommand\theHsection{S\arabic{section}}
\renewcommand\thetable{S\arabic{table}}

\setcounter{figure}{0}
\setcounter{section}{0}
\setcounter{equation}{0}
\setcounter{page}{1}
\setcounter{table}{0}

\begin{table}[h]
\centering
\setlength{\tabcolsep}{4pt}
\renewcommand{\arraystretch}{1.15}
\footnotesize
\begin{tabularx}{\textwidth}{r l l X X}
\toprule
\# & Model & Manufacturer & Target gas(es) & Detection range \\
\midrule
1  & MG812        & Winsen   & CO$_2$ & 350--10{,}000\,ppm \\
2  & WSP2110      & Winsen   & VOCs (toluene, benzene, formaldehyde, alcohol, acetone) & 1--50\,ppm \\
3  & TGS1820      & Figaro   & Acetone & 1--20\,ppm \\
4  & TGS2600      & Figaro   & Air contaminants (e.g., H$_2$, ethanol) & 1--30\,ppm (as H$_2$) \\
5  & TGS2602      & Figaro   & Air contaminants (VOCs, NH$_3$, H$_2$S, etc.) & 1--30\,ppm (as EtOH) \\
6  & TGS2603      & Figaro   & Odorants/air contaminants (amines, sulfur compounds) & 1--10\,ppm (as EtOH) \\
7  & TGS2610-D00  & Figaro   & LP gas (propane, butane) & 1--25\%\,LEL \\
8  & TGS2611-E00  & Figaro   & Methane (CH$_4$) & 1--25\%\,LEL \\
9  & TGS2612-D00  & Figaro   & Methane, propane, iso-butane & 1--25\%\,LEL (each) \\
10 & TGS2616-C00  & Figaro   & Hydrogen (H$_2$) & 30--3{,}000\,ppm \\
11 & TGS2620      & Figaro   & Organic solvent vapors (alcohols, etc.) & 50--5{,}000\,ppm (EtOH) \\
12 & TGS2630      & Figaro   & Refrigerant gases & 1{,}000--10{,}000\,ppm \\
13 & TGS3830      & Figaro   & Halocarbon refrigerants (e.g., R-134a) & 5--100\,ppm \\
14 & TGS3870-B00  & Figaro   & Methane + carbon monoxide & CH$_4$: 500--12{,}500\,ppm; CO: 50--1{,}000\,ppm \\
15 & SP3S-AQ2     & FIS Inc  & VOCs / air quality (MOS) & n/s\footnotemark[1] \\
16 & GSBT11       & Ogam Inc & VOCs, hydrocarbons, smoke, organic compounds (e.g., formaldehyde, toluene) & n/s\footnotemark[2] \\
\bottomrule
\end{tabularx}
\caption{Sensor target gases and detection ranges (typical values from manufacturers).}
\label{tab:sensor_array}
\end{table}

\footnotetext[1]{FIS SP3S-AQ2 datasheet provides qualitative VOC sensitivity/calibration info but no ppm detection range.}
\footnotetext[2]{OGAM GSBT11 materials list target VOCs; no numeric detection range is provided.}

% \begin{figure}[h]
% -    \centering
%     \includegraphics[width=0.8
%     \linewidth]{figs/pcb_render.pdf}
%     \vspace{5pt}
%     \caption{\textbf{
%     }}
%     \label{fig:extended_pcb}
% \end{figure}

\begin{figure}[h]
    \centering
    \includegraphics[width=\linewidth]{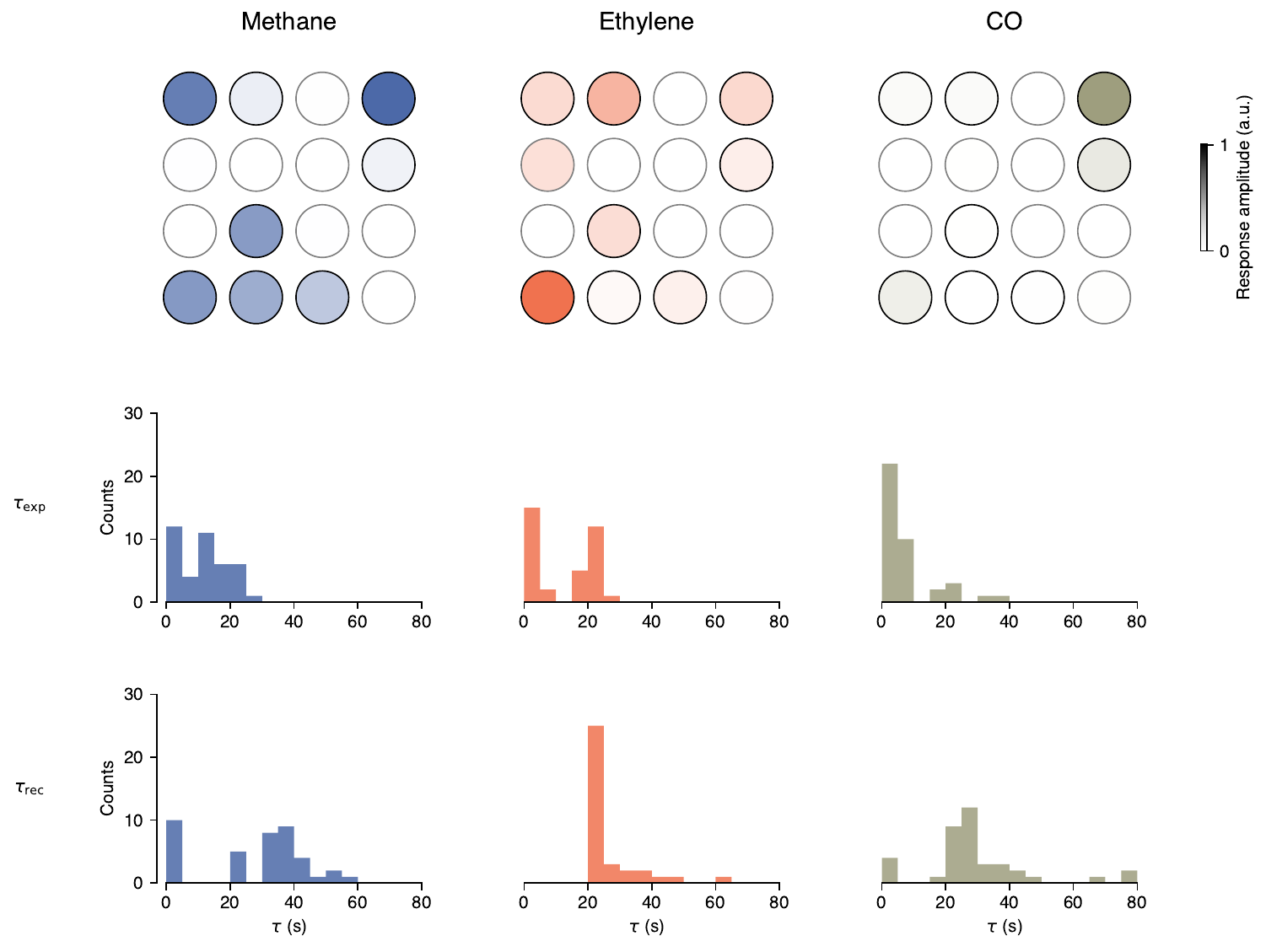}
    \vspace{5pt}
    \caption{\textbf{Sensor array characterisation.}
    Response of the sensor array to single-gas baseline–exposure–recovery cycles for the three analytes (columns: methane, ethylene, CO). Top row: spatial maps of the mean response amplitude over the physical $4\times4$ sensor layout, averaged over five repetitions and min–max normalised per analyte to $[0,1]$ (arbitrary units; colour bar). Marker outlines indicate whether a sensor is used in the classification experiments — black for the eight most-responsive sensors retained by the reduced data loader, grey for the excluded ones. Middle and bottom rows: distributions of the exposure ($\tau_{\mathrm{exp}}$) and recovery ($\tau_{\mathrm{rec}}$) time constants, obtained from single-exponential fits to the rising and decaying transients and pooled over the five repetitions and the eight retained sensors. The time constants lie predominantly in the $20$–$40,\mathrm{s}$ range, consistent with the sensor response times quoted in the data-acquisition protocol.
    }
    \label{fig:extended_sensor_characterization}
\end{figure}

% \begin{figure}[h]
%     \centering
%     \includegraphics[width=\linewidth]{figs/sensor_signal_with_noise.pdf}
%     \vspace{5pt}
%     \caption{\textbf{
%     }}
%     \label{fig:extended_sensor_signal_noise}
% \end{figure}

\end{document}

%% file: glossary.tex
\PassOptionsToPackage{acronym}{glossaries}

\glsdisablehyper

\newcommand*\myglsentry[1]{%
   \glsentrylong{#1}%
}

\newacronym[longplural={Frames per Second}]{fpsLabel}{FPS}{Frame per Second}
\newacronym[longplural={Tables of Contents}]{tocLabel}{TOC}{Table of Content}
\newacronym{act}{ACT}{Asynchronous Circuit Toolkit}
\newacronym{adc}{ADC}{Analog Digital Converter}
\newacronym{aer}{AER}{Address Event Representation}
\newacronym{afe}{AFE}{Analog FrontEnd}
\newacronym{ai}{AI}{Artificial Intelligence}
\newacronym{ampa}{AMPA}{$\alpha$-amino-3-hydroxy-5-methyl-4-isoxazolepropionic acid}
\newacronym{ams}{AMS}{Analog Mixed-Signal}
\newacronym{al}{AL}{Antennal Lobe}
\newacronym{api}{API}{Application Programming Interface}
\newacronym{auc}{AUC}{Area Under the Curve}
\newacronym{bap}{BAP}{back-propagating action potential}
\newacronym{bcall}{BCaLL}{Bi-Stable Calcium based Local Learning}
\newacronym{bd}{BD}{Bundled Data}
\newacronym{beol}{BEOL}{Back-End Of Line}
\newacronym{bjt}{BJT}{Bipolar Junction Transistor}
\newacronym{bnc}{BNC}{Bayonet-Neill-Concelman}
\newacronym{bptt}{BPTT}{Back Propagation Through Time}
\newacronym{bp}{BP}{Back Propagation}
\newacronym{camp}{cAMP}{cyclic Adenosine Mono Phosphate}
\newacronym{cam}{CAM}{Content Addressable Memory}
\newacronym{cco}{CCO}{Current Controlled Oscillator}
\newacronym{cc-by}{CC-BY}{Creative Commons Attribution}
\newacronym{chp}{CHP}{Communicating Hardware Processes}
\newacronym{cl}{CL}{Closed-Loop}
\newacronym{clt}{CLT}{Central Limit Theorem}
\newacronym{cm}{CM}{Continuum Mechanics}
\newacronym{cpu}{CPU}{Central Process Unit}
\newacronym{csv}{CSV}{Comma-Separated Values}
\newacronym{ctat}{CTAT}{Complementary To Absolute Temperature}
\newacronym{cuba}{CUBA}{Current Based}
\newacronym{dac}{DAC}{Digital to Analog Converter}
\newacronym{dcn}{DCN}{Dorsal Column Nuclei}
\newacronym{dc}{DC}{Direct Current}
\newacronym{dpi}{DPI}{Differential Pair Integrator}
\newacronym{dpss}{DPSS}{Dendritic Prediction of Somatic Spiking}
\newacronym{drg}{DRG}{Dorsal Root Ganglion}
\newacronym{dr}{DR}{Dual Rail}
\newacronym{dsnn}{D-SNN}{Deep Spiking Neural Network}
\newacronym{dsp}{DSP}{Digital Signal Processing}
\newacronym{ds}{DS}{Delay Sensitive}
\newacronym{dvs}{DVS}{Dynamic Vision Sensor}
\newacronym{eda}{EDA}{Electronic Design Automation}
\newacronym{elm}{ELM}{Extreme Learning Machine}
\newacronym{esn}{ESN}{Echo State Network}
\newacronym{fac}{FAC}{Facilitatory Trace}
\newacronym{fecap}{FeCap}{Ferroelectric Capacitor}
\newacronym{fet}{FET}{Field Effect Transistor}
\newacronym{fft}{FFT}{Fast Fourier Transform}
\newacronym{fifo}{FIFO}{First In First Out Register}
\newacronym{fi}{FI}{Frequency vs. Current}
\newacronym{fm}{FM}{Frequency Modulated}
\newacronym{fo}{FO}{First Order}
\newacronym{fpga}{FPGA}{Field Programmable Gate Array}
\newacronym{fsm}{FSM}{Finite State Machine}
\newacronym{ftj}{FTJ}{Ferroelectric Tunnel Junction}
\newacronym{ft}{FT}{Fourier transform}
\newacronym{gd}{GD}{Gradient Descent}
\newacronym{gmm}{GMM}{Gaussian mixture models}
\newacronym{gpu}{GPU}{Graphic Process Unit}
\newacronym{hp}{HP}{High Pass Filter}
\newacronym{i2c}{I$^2$C}{Inter Integrated Circuit}
\newacronym{ic}{IC}{Integrated Circuit}
\newacronym{if}{IF}{Integrate and Fire}
\newacronym{imc}{IMC}{In-Memory Computing}
\newacronym{iot}{IoT}{Internet of Things}
\newacronym{io}{I/O}{Input and Output}
\newacronym{ip}{IP}{Intellectual Property Macro}
\newacronym{irh}{IRH}{Instantaneous Rate Histogram}
\newacronym{isi}{ISI}{Interspike Interval}
\newacronym{iwta}{I-WTA}{Inverted Winner-Take-All}
\newacronym{knn}{K-NN}{K-Nearest Neighbors}
\newacronym{kwta}{k-WTA}{k-Winner-Take-All}
\newacronym{lcadc}{LC-ADC}{Level Crossing Analog Digital Converter}
\newacronym{lda}{LDA}{Linear Discriminant Analysis}
\newacronym{lif}{LIF}{Leaky \myglsentry{if}}
\newacronym{llm}{LLM}{Large Language Model}
\newacronym{lpf}{LPF}{Low Pass Filter}
\newacronym{lsh}{LSH}{Locality Sensitive Hashing}
\newacronym{lstm}{LSTM}{Long Short-Term Memory}
\newacronym{ltp}{LTP}{Long Term Plasticity}
\newacronym{lut}{LUT}{Look Up Table}
\newacronym{lvds}{LVDS}{Low Voltage Differential Signalling}
\newacronym{mac}{MAC}{Multiply Accumulate}
\newacronym{mb}{MB}{Mushroom Body}
\newacronym{mems}{MEMS}{Micro Electro-Mechanical System}
\newacronym{mfc}{MFC}{Mass Flow Controller}
\newacronym{mim}{MIM}{Metal Insulator Metal}
\newacronym{mcu}{MCU}{Microcontroller Unit}
\newacronym{mlm}{MLM}{Multi Layer Mask}
\newacronym{mlp}{MLP}{Multi Layer Perceptron}
\newacronym{ml}{ML}{Machine Learning}
\newacronym{mnist}{MNIST}{modified National Institute of Standards and Technology database}
\newacronym{mom}{MOM}{Metal Oxide Metal}
\newacronym{mos}{MOS}{Metal Oxide Semiconductor}
\newacronym{mox}{MOX}{Metal Oxide}
\newacronym{mtj}{MTJ}{Magnetic Tunnel Junction}
\newacronym{mvm}{MVM}{Matrix-Vector Multiplication}
\newacronym{nas}{NAS}{Neuromorphic Auditory Sensor}
\newacronym{nda}{NDA}{Non Disclosure Agreement}
\newacronym{nmda}{NMDA}{N-methyl-D-aspartate}
\newacronym{nni}{NNI}{Neural Network Intelligence}
\newacronym{nn}{NN}{Neural Network}
\newacronym{noc}{NoC}{Network on Chip}
\newacronym{nvm}{NVM}{Non-Volatile Memory}
\newacronym{oa}{OA}{OpenAccess}
\newacronym{opamp}{OPAMP}{Operational Amplifier}
\newacronym{orn}{ORN}{Olfactory Receptor Neuron}
\newacronym{ota}{OTA}{Operational Transconductance Amplifier}
\newacronym{pca}{PCA}{Principal Component Analysis}
\newacronym{pcb}{PCB}{Printed Circuit Board}
\newacronym{pcfb}{PCFB}{Pre-Charge Full Buffer}
\newacronym{pchb}{PCHB}{Pre-Charge Half Buffer}
\newacronym{pcm}{PCM}{Phase Change Material}
\newacronym{pc}{PC}{Pacini}
\newacronym{pdk}{PDK}{Process Development Kit}
\newacronym{pd}{PD}{Phase Detector}
\newacronym{pll}{PLL}{Phase-Locked Loop}
\newacronym{posfet}{POS-FET}{Piezoelectric Oxide Semiconductor Field Effect Transistor}
\newacronym{prs}{PRS}{Production Rule Set}
\newacronym{pr}{P\&R}{Place and Route}
\newacronym{psc}{PSC}{Post Synaptic Current}
\newacronym{ptat}{PTAT}{Proportional To Absolute Temperature}
\newacronym{pv}{PV}{Parietal Ventral Area}
\newacronym{pwm}{PWM}{Pulse Width Modulated}
\newacronym{qdi}{QDI}{Quasi Delay Insensitive}
\newacronym{qfp}{QFP}{Quad-Flat Package}
\newacronym{ra1}{RA}{Rapid-Adapting I}
\newacronym{ra2}{RA2}{Rapid-Adapting II}
\newacronym{rafr}{RaFr}{Radio Frequency}
\newacronym{ram}{RAM}{Random Access Memory}
\newacronym{rbssg}{RBSSG}{Reverse Bitwise Synthetic Spike Generator}
\newacronym{rf}{RF}{Receptive Field}
\newacronym{rl}{RL}{Reinforcement Learning}
\newacronym{rnn}{RNN}{Recurrent Neural Network}
\newacronym{roi}{ROI}{Region Of Interest}
\newacronym{roc}{ROC}{Receiver Operating Characteristic}
\newacronym{rpe}{RPE}{Reward Prediction Error}
\newacronym{s1}{S1}{Somatosensory Primary Cortex}
\newacronym{s2}{S2}{Somatosensory Secondary Cortex}
\newacronym{sa1}{SA1}{Slow-Adapting I}
\newacronym{sa2}{SA2}{Slow-Adapting II}
\newacronym{sdsp}{SDSP}{Spike-Driven Synaptic Plasticity}
\newacronym{sfd}{SFD}{Spike Frequency Divider}
\newacronym{shf}{S-HF}{Spike-based Hold \& Fire}
\newacronym{sig}{S-IG}{Spike-based Integrate \& Fire}
\newacronym{simd}{SIMD}{Single Instruction, Multiple Data}
\newacronym{slpf}{S-LPF}{Spike-based Low Pass Filter}
\newacronym{smd}{SMD}{Surface Mount Device}
\newacronym{smu}{SMU}{Source Measurement Unit}
\newacronym{soc}{SoC}{System on Chip}
\newacronym{sota}{SOTA}{State-Of-The-Art}
\newacronym{so}{SO}{Second Order}
\newacronym{spice}{SPICE}{Simulation Program with Integrated Circuit Emphasis}
\newacronym{spi}{SPI}{Serial Peripheral Interface}
\newacronym{spll}{sPLL}{Spiking Phase-Locked Loop}
\newacronym{src}{SRC}{Sparse Representation Classifier}
\newacronym{srdp}{SRDP}{Spike-Rate-Dependent Plasticity}
\newacronym{srm}{SRM}{Simple Response Model}
\newacronym{stdp}{STDP}{Spike Timing Dependent Plasticity}
\newacronym{stp}{STP}{Short Term Plasticity}
\newacronym{svm}{SVM}{Support Vector Machine}
\newacronym{syn}{SYN}{Synaptic Trace}
\newacronym{tc}{TC}{Temporal Contrast}
\newacronym{tde}{TDE}{Time Difference Encoder}
\newacronym{tpu}{TPU}{Tensor Processing Unit}
\newacronym{trg}{TRG}{Trigger Trace}
\newacronym{ttfs}{TTFS}{Time to first spike}
\newacronym{uart}{UART}{Universal Asyncronous Receiver-Transmitter}
\newacronym{uc}{$\mu$C}{Microcontroller}
\newacronym{vco}{VCO}{Voltage Controlled Oscillator}
\newacronym{vpd}{VP-d}{Victor-Purpura Distance}
\newacronym{wta}{WTA}{Winner Take All}
\newacronym{hd}{HD}{High-Dimensional}
\newacronym{rbf}{RBF}{Radial Basis Function}
\newacronym{kc}{KC}{Kenyon Cell}
\newacronym{pn}{PN}{Projection Neuron}
\newacronym{mbon}{MBON}{Mushroom Body Output Neuron}
\newacronym{vsa}{VSA}{Vector Symbolic Architecture}
\newacronym{ln}{LN}{Lateral Neuron}
\newacronym{bwta}{bWTA}{block-Winner-Take-All}
\newacronym{cdf}{CDF}{Cumulative Distribution Function}

\newacronym{alif}{ALIF}{Adaptive \myglsentry{lif}}
\newacronym{ann}{ANN}{Artificial \myglsentry{nn}}
\newacronym{asic}{ASIC}{Application Specific \myglsentry{ic}}
\newacronym{cmos}{CMOS}{Complementary \myglsentry{mos}}
\newacronym{cnn}{CNN}{Convolutional \myglsentry{nn}}
\newacronym{cstdp}{C-STDP}{Calcium \myglsentry{stdp}}
\newacronym{ddpi}{DDPI}{Double \myglsentry{dpi}}
\newacronym{dnn}{DNN}{Deep \myglsentry{nn}}
\newacronym{eai}{Edge-AI}{Edge \myglsentry{ai}}
\newacronym{enn}{ENN}{event-based \myglsentry{nn}}
\newacronym{epsc}{EPSC}{Excitatory \myglsentry{psc}}
\newacronym{exlif}{ExLIF}{Exponential \myglsentry{lif}}
\newacronym{fefet}{FeFET}{Ferroelectric \myglsentry{fet}}
\newacronym{gpgpu}{GPGPU}{General-Purpose computing on \myglsentry{gpu}}
\newacronym{ipsc}{IPSC}{Inhibitory \myglsentry{psc}}
\newacronym{moscap}{MOSCAP}{\myglsentry{mos} Capacitor}
\newacronym{mosfet}{MOSFET}{\myglsentry{mos} \myglsentry{fet}}
\newacronym{nmnist}{N-MNIST}{Neuromorphic \myglsentry{mnist}}
\newacronym{nmos}{nMOS}{n-type \myglsentry{mos}}
\newacronym{oxram}{OxRAM}{Oxide-based resistive \myglsentry{ram}}
\newacronym{pka}{PKA}{\myglsentry{camp}-depended Protein Kinase}
\newacronym{pmos}{pMOS}{p-type \myglsentry{mos}}
\newacronym{rram}{RRAM}{Resistive \myglsentry{ram}}
\newacronym{sadc}{sADC}{spiking \myglsentry{adc}}
\newacronym{scnn}{sCNN}{spiking \myglsentry{cnn}}
\newacronym{snn}{SNN}{Spiking \myglsentry{nn}}
\newacronym{sodpi}{SoDPI}{Second-order \myglsentry{dpi}}
\newacronym{sram}{SRAM}{Static \myglsentry{ram}}
\newacronym{tstdp}{T-STDP}{Triplet \myglsentry{stdp}}

\newacronym{adexlif}{AdExLIF}{Adaptive \myglsentry{exlif}}
\newacronym{nfet}{n-FET}{\myglsentry{nmos} \myglsentry{fet}}
\newacronym{pfet}{p-FET}{\myglsentry{pmos} \myglsentry{fet}}

\newacronym{sma}{SMA}{SubMiniature version A}
\newacronym{wrota}{WR-OTA}{Wide Range \gls{ota}}
\newacronym{sstdp}{S-STDP}{Stochastic \myglsentry{stdp}}
\newacronym{neuop}{NeuOp}{Neuron spike Operation}
\newacronym{synop}{SynOp}{Synaptic Operation}
\newacronym{fdsoi}{FDSOI}{Fully Depleted Silicon On Insulator}
\newacronym{qif}{QIF}{Quadratic \myglsentry{if}}
\newacronym{ff}{FF}{edge triggered Flip-Flop}
\newacronym{lrs}{LRS}{Low Resistance State}
\newacronym{hrs}{HRS}{High Resistance State}

\makeglossaries